\PassOptionsToPackage{dvipsnames}{xcolor}
\documentclass[colorlinks=true, linkcolor=blue, citecolor=blue, filecolor=blue, urlcolor=blue]{aa}

\usepackage{lipsum}
\usepackage{physics}
\usepackage{multirow}
\usepackage{xspace}
\usepackage{natbib}
\usepackage{fontawesome5}
\usepackage{wrapfig}
\usepackage[figuresright]{rotating}
\usepackage{fontawesome5}
\usepackage{listings}
\usepackage{xcolor} 
\usepackage[varg]{txfonts}

\usepackage{float}
\usepackage{graphicx}
\usepackage{txfonts}
\usepackage{orcidlink}          

\usepackage[hang,flushmargin]{footmisc} 

\newcommand{\Msun}{\,M$_{\odot}$} 
\newcommand{\Rsun}{\,R$_{\odot}$} 
\newcommand{\cpd}{\,d$^{-1}$} 

\newcommand{\bcep}{$\beta$\,Cep}
\newcommand{\MESA}{\textsc{mesa}}

\newcommand{\BV}{Brunt-V\"{a}is\"{a}l\"{a} }

\newcommand\T{\rule{0pt}{2.6ex}}       
\newcommand\B{\rule[-1.2ex]{0pt}{0pt}} 

\usepackage{color}
\definecolor{green(new)}{RGB}{50, 200, 60}

\definecolor{oror}{RGB}{0,150,0}

\usepackage{amsmath}

\usepackage[colorinlistoftodos]{todonotes}

\let\Oldtodo\todo
\renewcommand{\todo}[1]{\Oldtodo[inline, nolist, noinlinepar]{\textbf{TODO:} #1}}

\begin{document}

\title{The evolutionary and asteroseismic imprints of mass accretion \\ The $10\,\mathrm{M_\odot}$\ $\beta\ \mathrm{Cep}$ case study}

\author{A. Miszuda \inst{1} \orcidlink{0000-0002-9382-2542}
    \and Z. Guo \inst{2} \orcidlink{0000-0002-0951-2171}
    \and R. H. D. Townsend \inst{3} \orcidlink{0000-0002-2522-8605}
    }

\institute{Nicolaus Copernicus Astronomical Centre, Polish Academy of Sciences, Bartycka 18, PL-00-716 Warsaw,  Poland \\ \email{amiszuda@camk.edu.pl, amiszuda.astro@gmail.com}
    \and Institute of Astronomy, KU Leuven, Celestijnenlaan 200D, B-3001 Leuven, Belgium
    \and Department of Astronomy, University of Wisconsin-Madison, 475 N Charter St, Madison, WI 53706, USA
}

\date{Received \today{}}

\abstract{
We investigate the structural and asteroseismic consequences of mass accretion in massive stars within close binary systems. Using \MESA, we model the evolution of the $10\ \rm M_{\odot}$ accretor through and after a Roche lobe overflow phase. In addition to changing the surface composition of the star, mass accretion also significantly modifies the internal structure by expanding the convective core and altering chemical stratification near the core-envelope boundary. This partial core rejuvenation creates a distinct mean molecular weight gradient and leaves a persistent local density modulation.

In the late stages of mass transfer, changes in density and sound-speed profiles become apparent and influence stellar oscillations. We analyse the asteroseismic properties of the post-mass transfer models compared to single stars of the same mass and central hydrogen abundance. In the gravity mode regime, the altered Brunt-V\"{a}is\"{a}l\"{a} frequency leads to period spacing patterns with larger amplitudes and phase shifts. For low- and intermediate-order pressure modes, we find systematic frequency deviations linked to changes in the sound-speed profile. Weight function analyses confirm that these differences arise primarily from structural modifications near the convective core boundary.

Furthermore, small frequency separations, sensitive to localized sound-speed gradients, reveal periodic variations attributable to the density discontinuity at the convective core edge. The accretor exhibits a larger sound-speed gradient integral and a longer acoustic radius ratio compared to the single star, consistent with its expanded core.

Our results demonstrate that mass accretion imprints measurable asteroseismic signatures on both gravity and pressure modes. These signatures provide powerful diagnostics for identifying post-interaction stars and for refining stellar age and structure estimates in binary systems.
}

\maketitle

\section{Introduction} \label{sec:introduction}

Binary stars constitute a significant fraction of the stellar population in the Galaxy. Observations indicate that over half of massive stars are born in binary or multiple systems \citep{Duchene2013,2012Sci...337..444S}, with possible interactions between components shaping their subsequent evolution. One of the key processes in such systems is mass transfer, which occurs when one star fills its Roche lobe and begins transferring material to its companion. This mass exchange profoundly alters both stars’ evolution, often leading to rejuvenation, expansion and spin-up of one component or stripping of the other \citep{1992ApJ...391..246P,deMink2013}.

Mass transfer in close binaries leads to a range of structural and orbital consequences. As mass is accreted by the initially less massive star, the binary orbit responds according to angular momentum conservation. The increasing mass of the accretor and corresponding mass ratio reversal typically result in orbital expansion during the later stages of the interaction \citep{1997A&A...327..620S,Renzo2023}. However, initially the transfer of mass from the more massive donor causes the orbit to shrink \citep[eg.,][]{Renzo2019}. At the same time, the internal structure of the accreting star undergoes significant changes \citep{Renzo2021}. As material is deposited onto the surface, it is gradually incorporated into the stellar interior through thermohaline, convective and rotational mixing processes, particularly near the core-envelope boundary \citep{Neo1977}. This mixing can supply the convective core with fresh hydrogen, effectively rejuvenating the star. The phenomenon of core rejuvenation leads to an increase in convective core mass and alters the subsequent nuclear burning history of the accretor. Moreover, the associated mixing processes can also alter the convective core size, internal chemical gradients, and rotational profile \citep{2017ApJS..230...15M,Renzo2021}. These changes in internal stratification have direct consequences for the star’s pulsational properties \citep{Guo2021,Miszuda2022,Wagg2024,2024A&A...690A..65H,2025A&A...698A..49H}.

Asteroseismology offers a unique window into stellar interiors by analysing oscillation modes sensitive to different regions within the star \citep[e.g.,][]{Aerts2010}. In particular, gravity modes ($g$ modes) probe stable, stratified regions, and specifically the chemical gradient and convective core boundaries, while pressure modes ($p$ modes) provide constraints on the sound-speed profile in outer layers \citep{Miglio2008}. In mass-accreting stars, modifications to the internal structure due to accretion are expected to leave characteristic signatures in both $g$ \citep{Wagg2024,2024A&A...690A..65H,2025A&A...698A..49H} and $p$ mode frequencies \citep{Miszuda2025CoSka..55c.196M}. In particular, the rejuvenation from mass transfer shifts the $p$ mode to higher frequencies \citep{Guo2017} and the $g$ mode period spacing to larger values \citep{Guo2019}. \citet{Arras2006} found that mass accretion can shift the edge of $g$ mode instability strip of the white dwarf star in Cataclysmic Variables.

Recent observational and theoretical works have identified such asteroseismic signatures in post-mass transfer binaries. Studies of slowly pulsating B-type stars revealed period spacing deviations consistent with altered chemical profiles from mass accretion \citep{Wagg2024}. Similarly, models of pre-main sequence stars indicate that accretion histories shape their pulsational behaviour and internal structure \citep{2022FrASS...9.4738Z}.

In this work, we investigate the impact of mass accretion on the internal mixing, structure, and pulsational properties of massive, 10\Msun\ accretor stars using detailed binary evolution modelling. By comparing accretor models with single-star counterparts, we identify key structural changes and their observational consequences, advancing the understanding of the role of mass transfer in stellar evolution.

\section{Methodology} 
\label{sec:methodology}

\subsection{Evolutionary calculations} 
\label{subsec:methodology:evolutionary_calculations}

We used the \textsc{mesa} code \citep[Modules for Experiments in Stellar Astrophysics,][version 23.05.1]{Paxton2011, Paxton2013, Paxton2015, Paxton2018, Paxton2019, Jermyn2023} to construct a toy model under the non-rotating approximation. 
The model was created both as a product of binary interactions (through mass gain) and as a single star, allowing us to compare the binary model to its isolated counterpart. The details on the variety of physical processes, which \MESA\ relies on, can be found in \autoref{appendix:mesa}.

In our evolutionary computations, we adopted the AGSS09 \citep{Asplund2009} initial chemical composition and used the OPAL opacity tables, supplemented by data from \cite{Ferguson2005} for lower temperatures. For high-temperature regimes, as well as hydrogen-poor or metal-rich conditions, we used CO enhanced tables. We assumed a metallicity of $Z = 0.014$ and an initial hydrogen abundance of $X_0 = 0.723$.

Convective instability in the models was treated using the Ledoux criterion, combined with the turbulent convection theory based on the \cite{Kuhfuss1986} model, employing a mixing-length parameter of $\alpha_{\rm MLT} = 0.5$. In regions that were stable according to the Ledoux criterion but unstable by the Schwarzschild criterion, we applied the semiconvective mixing following the formalism of \cite{Langer1985} with a scaling factor $\alpha_{\rm sc} = 0.1$. 
To address regions exhibiting an inversion in the mean molecular weight, such as those formed during mass accretion, we applied the thermohaline mixing using the formalism of \cite{Kippenhahn1980}, with an $\alpha_{\rm th} = 1$ coefficient. As our models did not include rotational mixing, we introduced a minimum diffusive mixing coefficient of $D = 100\  \rm cm^2\,s^{-1}$ to smooth out numerical noise or discontinuities in internal profiles, such as the Brunt-V\"{a}is\"{a}l\"{a} frequency or composition gradients. 
Additionally, we accounted for overshooting beyond the formal convective boundaries, using an overshooting parameter $f_{\rm ov}$, to capture turbulent motions extending into the radiative zone. We applied an exponential scheme of \cite{Herwig2000} on the top of the H-burning core with the value of $f_{\rm ov}=0.02$.

To ensure numerical convergence, we explored the numerical stability of our results against changes in both spatial and time step resolution, testing various configurations and ultimately adopting \texttt{mesh\_delta\_coeff~=~0.2} and \texttt{time\_delta\_coeff~=~0.5}. We have also tested the impact of the adopted value of the minimum diffusive mixing coefficient, ultimately adopting $D = 100\  \rm cm^2\,s^{-1}$. We selected this value after testing various options, specifically $D = 10$, 20, 50, 100 and 200, identifying it as the lowest value that effectively smoothed out numerical noise in the aforementioned profiles without erasing any significant physical structures.

Our initial binary model consists of a $10\,\rm M_{\odot}$ donor in a $3$-day circular orbit with a $7\,\rm M_{\odot}$ accretor. The initial orbital period was selected so that the donor fills its Roche lobe and initiates mass transfer while still on the main sequence. For simplicity, we assumed fully conservative mass transfer (MT), following the description of \cite{Kolb1990}. In this scenario, the initially less massive star accretes $3\,\rm M_{\odot}$, leading to a full mass ratio reversal. Once the accretor reached $10\,\rm M_{\odot}$, all binary interactions were stopped, and we assumed that the subsequent evolution proceeded as a single star, following the standard path until central hydrogen depletion. To simulate this behaviour in \MESA, we used the routine \textsc{detach\_binary} by Mathieu Renzo\footnote{https://github.com/MESAHub/mesa-contrib/tree/release/hooks/detach\_binary}.

For comparison purposes, we have also calculated single-star evolution for stars with masses from $7\,\rm M_{\odot}$ to $10\,\rm M_{\odot}$ with a $0.5\,\rm M_{\odot}$ step with the physics input as described above.

\subsection{Pulsational calculations} 
\label{subsec:methodology:pulsations}

To calculate pulsations using the previously described \MESA\ models, we employed the \textsc{gyre} code \citep[][version 7.1]{GYRE-Townsend2013,GYRE-Townsend2018,GYRE-Goldstein2020,GYRE-Sun2023}, in the adiabatic approximation. For radial pulsations ($\ell = 0$), we scanned the frequency range from 1 to 150\cpd\ using a linear grid. For higher mode numbers (non-radial pulsations, $\ell = 1, 2, 3, 4$), we applied an inverse grid sampling between 0.1 and 3.0\cpd\ and a linear grid from 3.0 to 150\cpd. The \textsc{gyre} calculations were performed for all models in our \MESA\ grid. 
These computations were applied to both single-star models and accretor models from binary evolution alike.

\section{Results} 
\label{sec:results}

\subsection{Binary evolution}

\begin{figure}[thbp]
    \includegraphics[width=\linewidth]{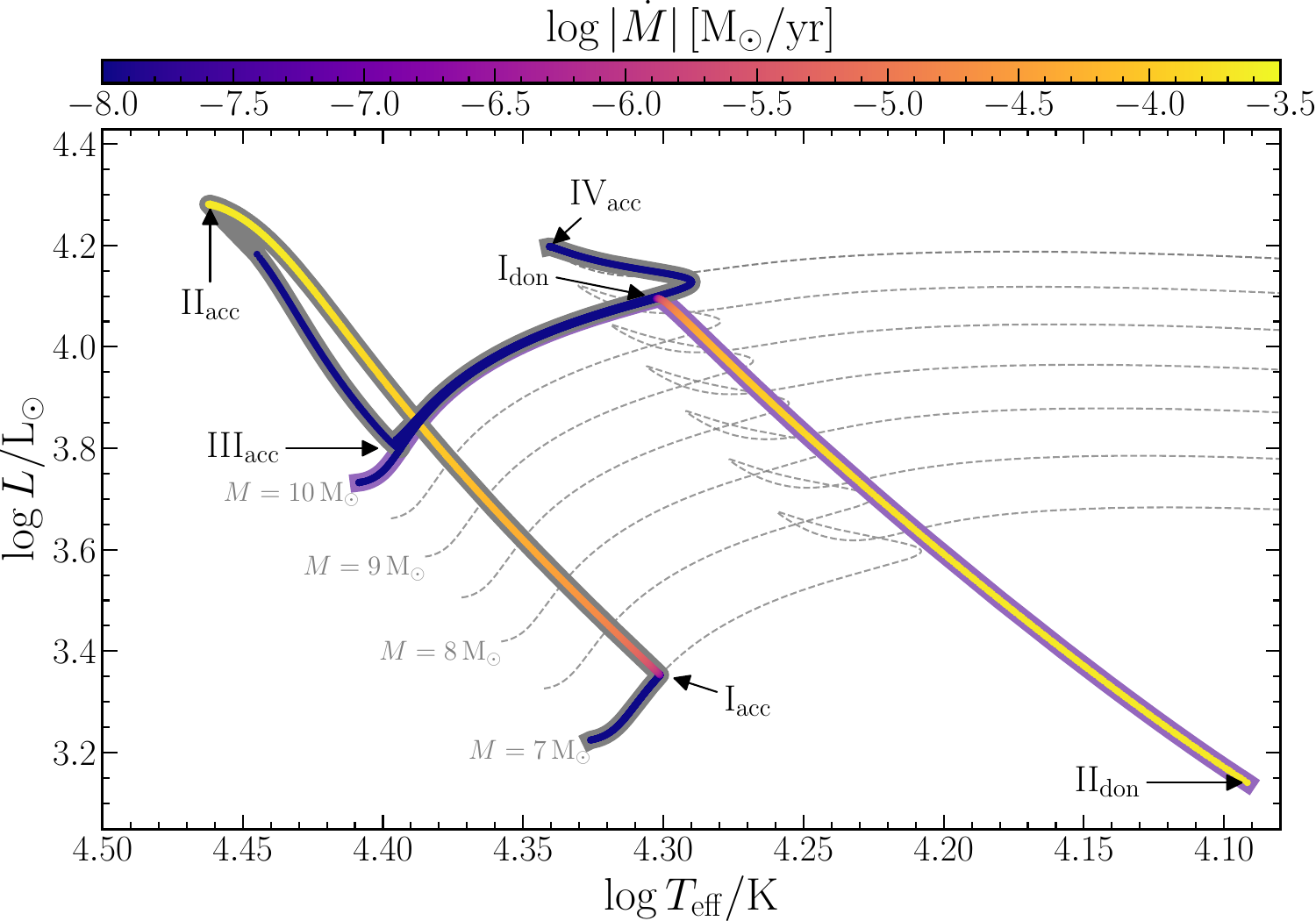}
    \caption{The Hertzsprung–Russell diagram with the evolutionary tracks of the binary system components. The donor star is shown with the purple line, and the accretor with the gray line, both plotted alongside single-star reference models for different masses (dashed lines). The mass transfer/accretion rate $\log \dot{|M|}$ is overplotted along the tracks using colour coding, as indicated by the colour bar. We mark key evolutionary changes: the donor fills its Roche lobe and begins mass transfer (I$_\mathrm{don}$), the accretor starts to gain mass and departs from its original track (I$_\mathrm{acc}$), full mass ratio reversal when the accretor reaches 10\,\Msun\ and the following detachment of the system (II$_\mathrm{acc}$ and II$_\mathrm{don}$), thermal readjustment of the accretor (III$_\mathrm{acc}$), and the accretor’s single-star evolution until the end of the main sequence (IV$_\mathrm{acc}$).}
\label{fig:HR}
\end{figure}

The early evolution of the system, initially containing 10\Msun\ and 7\Msun\ components, follows the single-star evolutionary paths, i.e. both stars detach from the Zero Age main sequence (ZAMS) and start synthesizing hydrogen in their cores. 
Their subsequent evolution is illustrated in the Hertzsprung–Russell (HR) diagram, in \autoref{fig:HR}.
As hydrogen burning proceeds in their convective cores, both components start to increase their radius and move through the HR diagram at different paces. By the time the 10\Msun\ star synthesized almost all central hydrogen into helium (approximately 10\% hydrogen left in the core), it has grown enough to fill its Roche lobe (at a radius of approximately 9.3\Rsun) and starts to transfer mass onto its companion ($\rm I_{don}$ in \autoref{fig:HR}). At this point the system is approximately 23.55\,Myr old and the components begin to diverge from their single track equivalents.

As the donor star begins to transfer mass through the inner Lagrange point $L_1$ onto its companion during a case A mass transfer episode \citep{Kippenhahn1967}, it remains in thermal equilibrium. The mass transfer proceeds stably, yet rapidly, with the rate reaching up to $\dot{M}\sim10^{-3.5}\ \mathrm{M_\odot\ yr^{-1}}$. The MT occurs on the nuclear timescale, and is primarily driven by the gradual expansion of the donor’s radius resulting from internal evolutionary changes.
As mass is transferred onto the initially less massive companion (starting from point $\rm I_{acc}$ in \autoref{fig:HR}), the accretor adjusts to the increasing mass by gradually expanding in size and evolving off its original evolutionary track. 
During MT, the accretor remains close to thermal equilibrium while accommodating the incoming material. As a result, it experiences a steady increase in mass, luminosity, effective temperature, and radius.
We allow the accretor to grow mass until it reaches 10\,\Msun, i.e. until full mass ratio reversal ($\rm II_{acc}$ in \autoref{fig:HR}). After this point, we artificially detach the system which allows the accretor to regain the thermal equilibrium in the Kelvin-Helmholtz time scale and follow the subsequent evolution of the accretor as a single 10\,\Msun\ star (starting from point $\rm III_{acc}$ in \autoref{fig:HR}) until the the end of the main sequence phase ($\rm IV_{acc}$ in \autoref{fig:HR}).

\subsection{General properties of the accretor model}

In the mass-accreting models, the incoming mass has a severe effect on the internal structure. As the transferred material binds to the star, the induced higher central temperature boosts the nuclear energy generation rate. As the core luminosity goes up, the radiative gradient $\nabla_{\rm rad}$ increases leading to a larger convective core region according to the convective instability criterion, as will be discussed in more detail later in this section.

\begin{figure}[tbp]
    \centering
    \includegraphics[width=\linewidth]{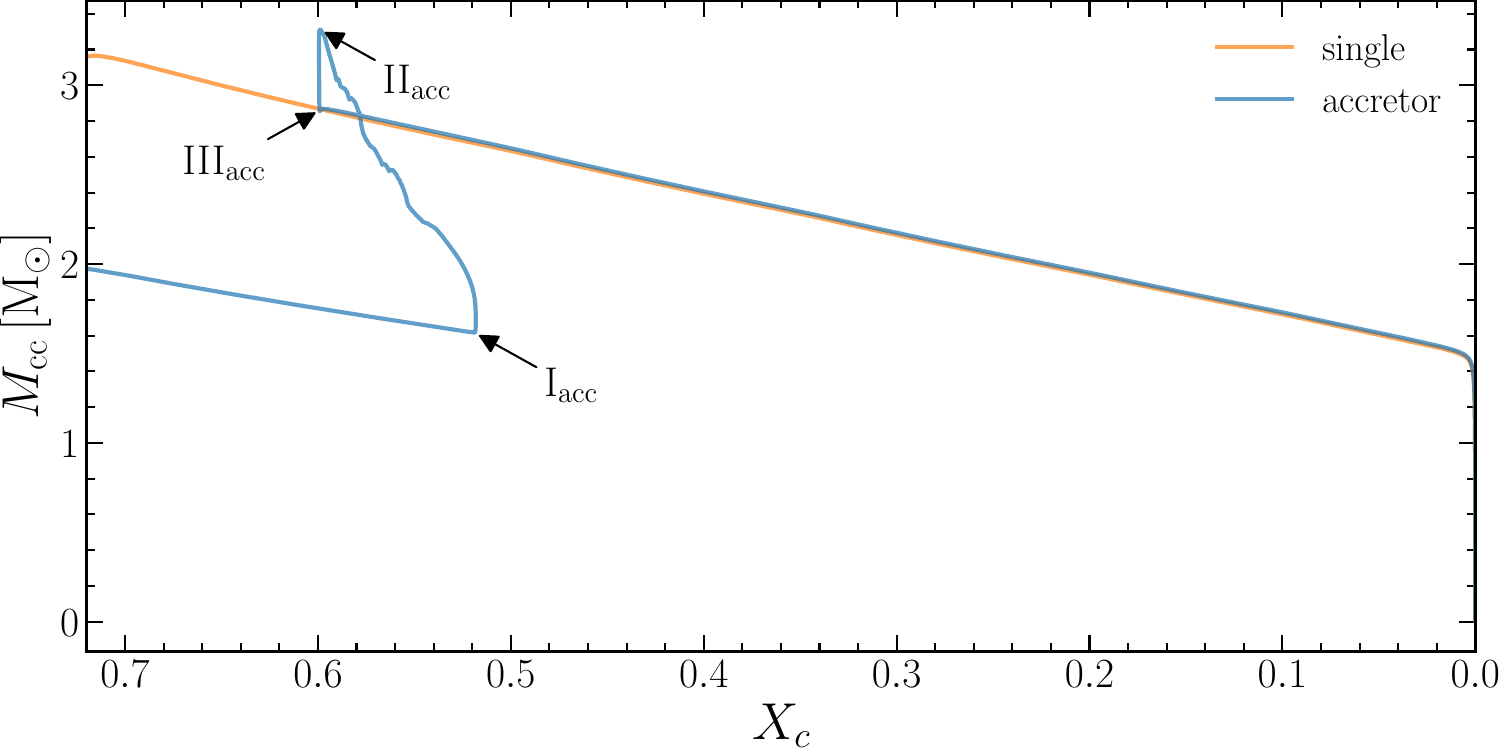}
    \caption{The evolution of mass of the convective core $M_\mathrm{cc}$ in a function of central hydrogen abundance $X_c$ for the single-star (orange lines) and the accretor (blue lines) models. We mark key evolutionary changes of the accretor according to \autoref{fig:HR}.}
\label{fig:convective_core_evol}
\end{figure}

The evolution of the convective core during this process is shown in \autoref{fig:convective_core_evol}, where the core mass $M_{\rm cc}$ is plotted against the central hydrogen abundance $X_c$. For the single-star model, the convective core steadily recedes from ZAMS to terminal age main sequence (TAMS). By contrast, the initially smaller convective core of the accretor follows a similar trend only until $X_c \approx 0.5$ ($\rm I_{acc}$, see \autoref{fig:HR}), after which it expands due to mass accretion. Within just 0.02 Myr, $M_{\rm cc}$ increases from 1.62\Msun\ to 3.3\Msun\ ($\rm II_{acc}$), reflecting the sudden influx of fresh fuel. Once mass transfer ceases, the star regains thermal equilibrium and the core mass stabilizes at $M_{\rm cc} = 2.87$\Msun\ with $X_c \approx 0.6$ ($\rm III_{acc}$). From that point onward, the evolution of the accretor’s convective core closely tracks that of the single-star counterpart, although it remains slightly more massive by about 1\%. This offset likely results from enhanced mixing across the core-envelope boundary \citep[CEB,][]{Neo1977,Renzo2023}, which altered the chemical profile and increased the core mass.

\begin{figure*}[tbp]
    \centering
    \begin{tabular}{cc}
        \includegraphics[width=0.45\textwidth]{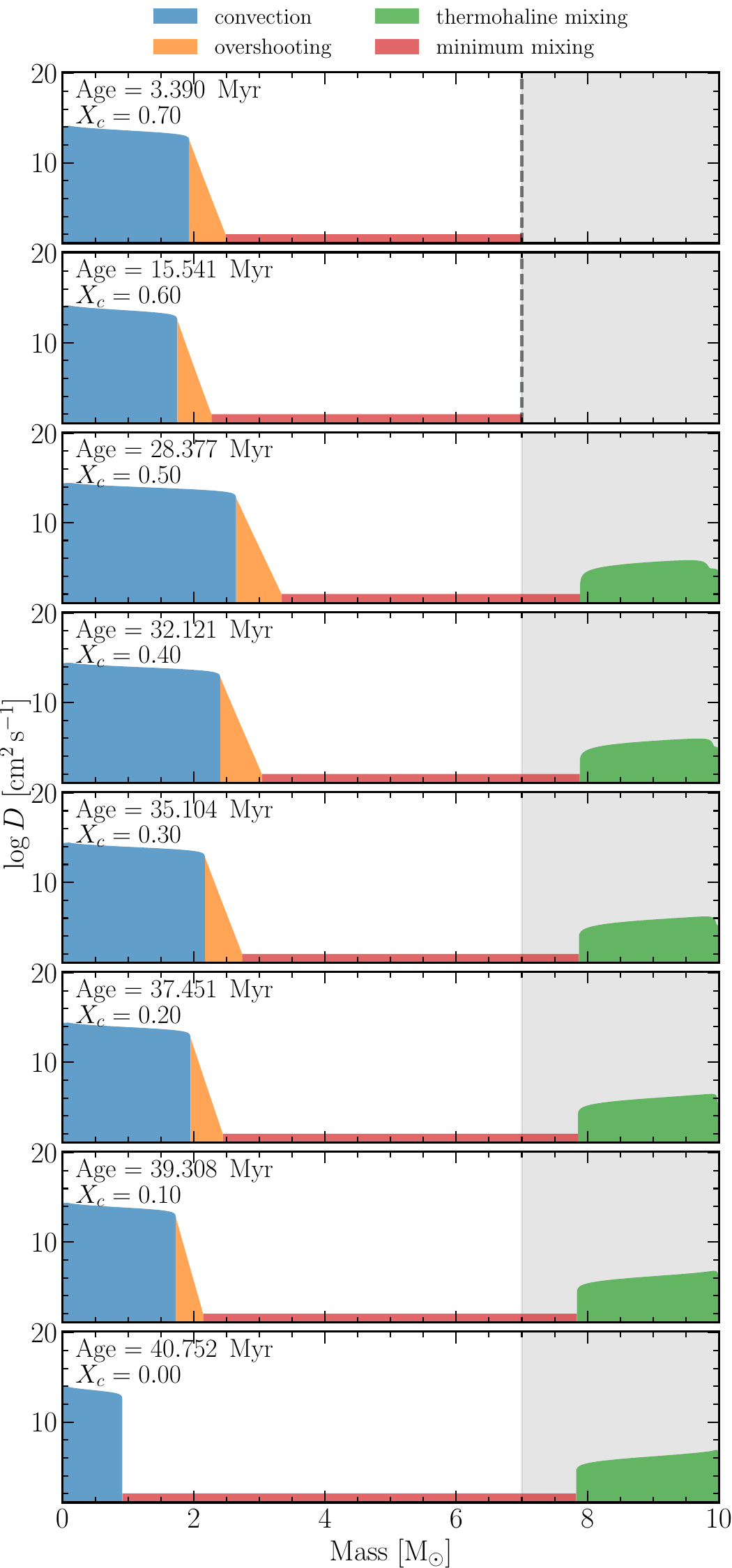} & 
        \includegraphics[width=0.45\textwidth]{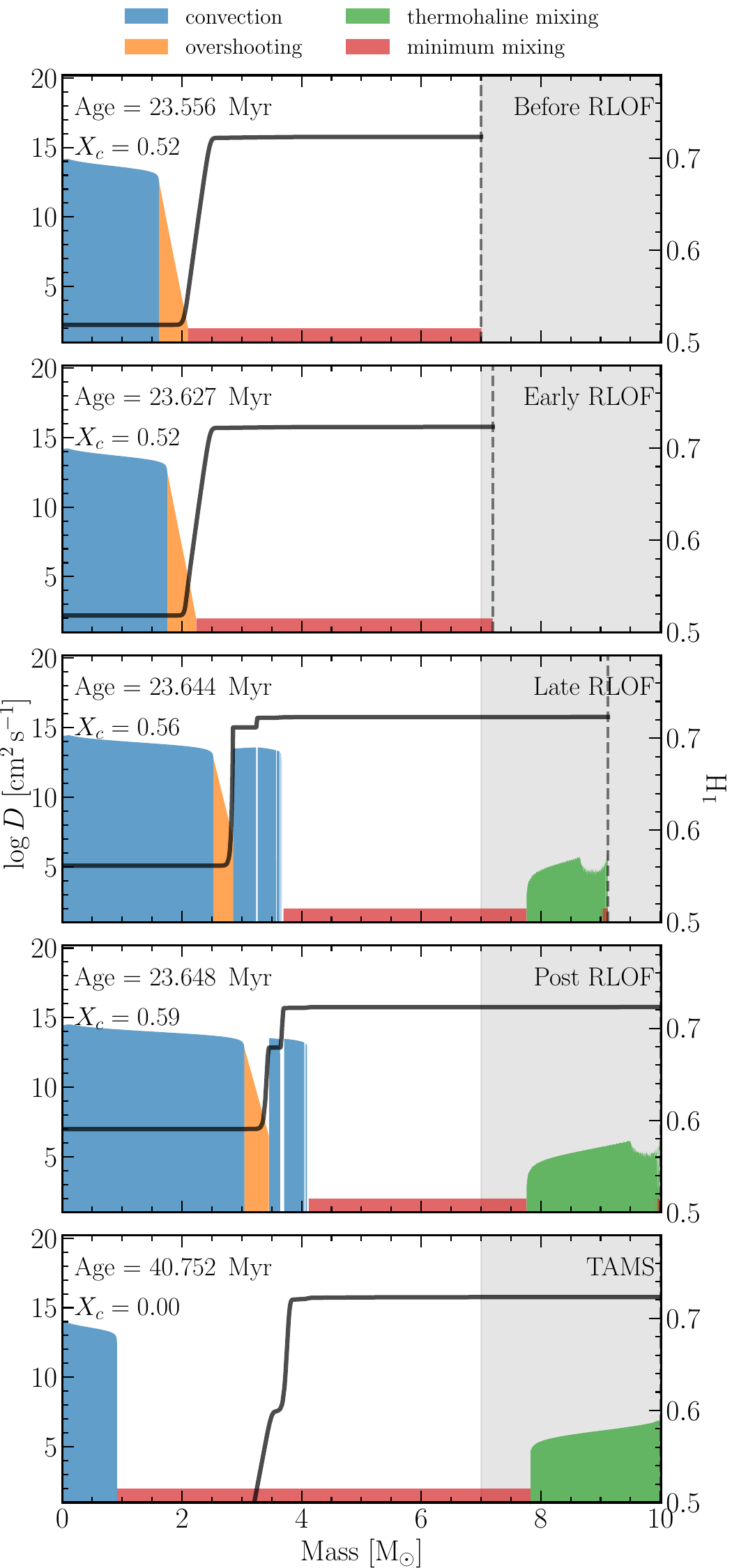} \\
    \end{tabular}
    \caption{The mixing profiles $D$ during the evolution of the accretor. Left: Evolution from $X_c = 0.7$ to 0.0 in steps of 0.1. Right: Profiles at selected evolutionary stages, before, during and after the RLOF. The shaded areas indicate the regions over which the mass varies, with the current mass indicated by dashed lines. The black lines, as denoted on the right-hand ordinates of the right panels show the corresponding profiles of hydrogen abundances.}
\label{fig:profiles}
\end{figure*}

The structural changes brought about by accretion are clearly visible in the evolution of internal mixing and hydrogen abundance profiles, shown in \autoref{fig:profiles}, based on the convection, overshoot, thermohaline and user-defined artificial mixing processes. 
Mass accretion is relatively short-lived and rapid and occurs between the left, second and third top panels. This figure allows one to clearly trace the evolution of the convective core boundary in mass coordinates, both during the standard core recession and the core rejuvenation phase.

\begin{figure}[thbp]
    \centering
    \includegraphics[width=\linewidth]{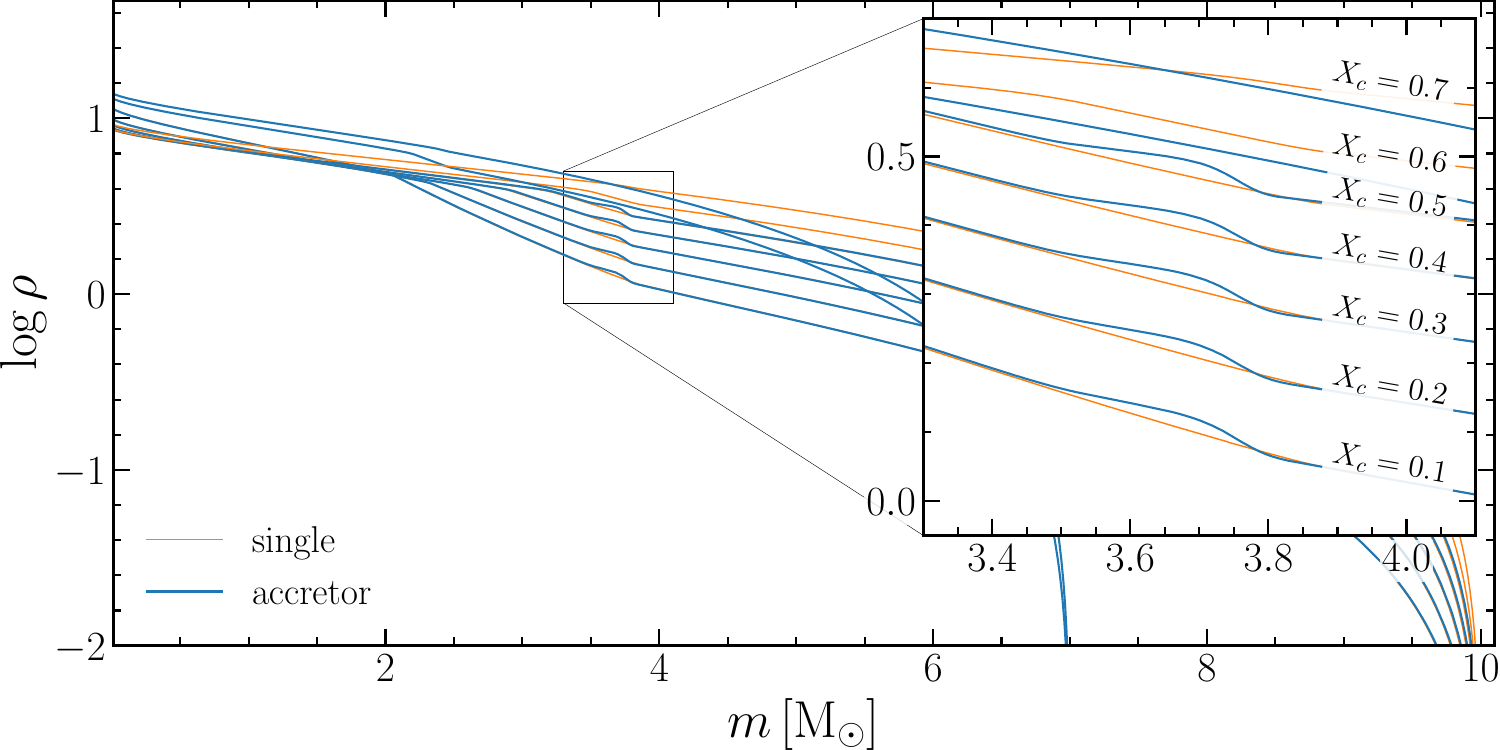}
    \caption{The density $\log \rho$ profiles at selected central hydrogen abundances $X_c$, for the single-star (orange lines) and the accretor (blue lines) models. The local bump discussed in the text is clearly visible in the binary models.}
\label{fig:rho}
\end{figure}

The effect of rejuvenation is twofold. 
As the freshly accreted material settles onto the stellar surface, an inversion in the mean molecular weight profile forms and gives rise to thermohaline mixing, which contributes to the internal redistribution of chemically stratified layers \citep{Kippenhahn1980,Stan2008,Wagg2024}.
The expansion of the convective core during accretion leads to the engulfment of layers with hydrogen gradient, left after the core recession, mixing hydrogen-rich material from the envelope into the core \citep{Neo1977}. At the same time, this process compresses the surrounding chemically stratified region, steepening the hydrogen gradient in the adjacent layers. As a result, the mean molecular weight $\mu$ drops sharply in this narrow zone, and an inflection in the density profile emerges. Specifically, a slightly increased density in the location of the previously engulfed regions is followed by a steeper decline due to drop in $\mu$, compared to the single models as shown in \autoref{fig:rho}. We refer to this feature as a local bump in the density profile. This bump reflects a deviation from the otherwise monotonic density gradient and emerges just above the growing convective core, in layers enriched in hydrogen as a signature of structural reconfiguration due to accretion. As the core expands during the mass accretion, it shifts this bump into higher layers, near the end of the hydrogen gradient region. This feature is stable, meaning that it persists at least until the end of the main sequence phase (see \autoref{fig:rho}). The presence of this bump is also consistent with earlier findings by \citet{Renzo2023} and \citet{Wagg2024}, though the models of the just mentioned authors describe a local \textit{dip} in the density profile.

A key consequence of the density bump is the development of off-centre convective zones (oCZs). These zones appear in the chemically stratified region above the convective core during the late and post-RLOF phases, as seen in the right panels of \autoref{fig:profiles}. Their occurrence has been reported by \citet{Renzo2021} and their physical origin was analysed by \cite{Miszuda2025_oCZs}.
The formation of oCZs is governed by the Ledoux criterion for convective instability. According to this condition, a region becomes unstable to convection if the actual temperature gradient exceeds the Ledoux gradient
\begin{equation}
\nabla > \nabla_{\rm L} = \nabla_{\rm ad} + \frac{\varphi}{\delta} \nabla_{\mu},
\end{equation}
where $\nabla$ is the temperature gradient, $\nabla_{\rm ad}$ is the adiabatic temperature gradient, and $\nabla_{\mu}$ is the gradient of the mean molecular weight:
\begin{equation}
\nabla = \frac{\mathrm{d} \ln T}{\mathrm{d} \ln P},\  \nabla_{\text{ad}} = \left( \frac{\partial \ln T}{\partial \ln P} \right)_{\text{ad}}\ \ \mathrm{and}\ \ \nabla_{\mu} = \frac{\mathrm{d} \ln \mu}{\mathrm{d} \ln P}\ .
\end{equation}
The thermodynamic derivatives 
\begin{equation}
\delta = -\left( \frac{\partial \ln \rho}{\partial \ln T} \right)_{P,\mu} \hspace{5pt} \mathrm{and}\hspace{10pt} \varphi = \left( \frac{\partial \ln \rho}{\partial \ln \mu} \right)_{P,T}
\end{equation}
are obtained from the equation of state \citep[e.g.,][]{Kippenhahn1990}. In radiative layers, where energy is transported by radiation, the actual temperature gradient $\nabla$ is equal to the radiative temperature gradient $\nabla_{\rm rad}$. Therefore, in these layers, the Ledoux instability criterion takes the form of
\begin{equation}
\nabla_{\rm rad} > \nabla_{\rm L}. 
\end{equation}
In the post-mass transfer models, the appearance of a local density bump leads to an increase in opacity, which enhances $\nabla_{\rm rad}$. If this exceeds the stabilising contribution from $\nabla_{\mu}$, the stability Ledoux criterion is violated and a convective zone forms away from the centre — giving rise to the oCZ.

\begin{figure}[tbp]
    \centering
    \includegraphics[width=\linewidth]{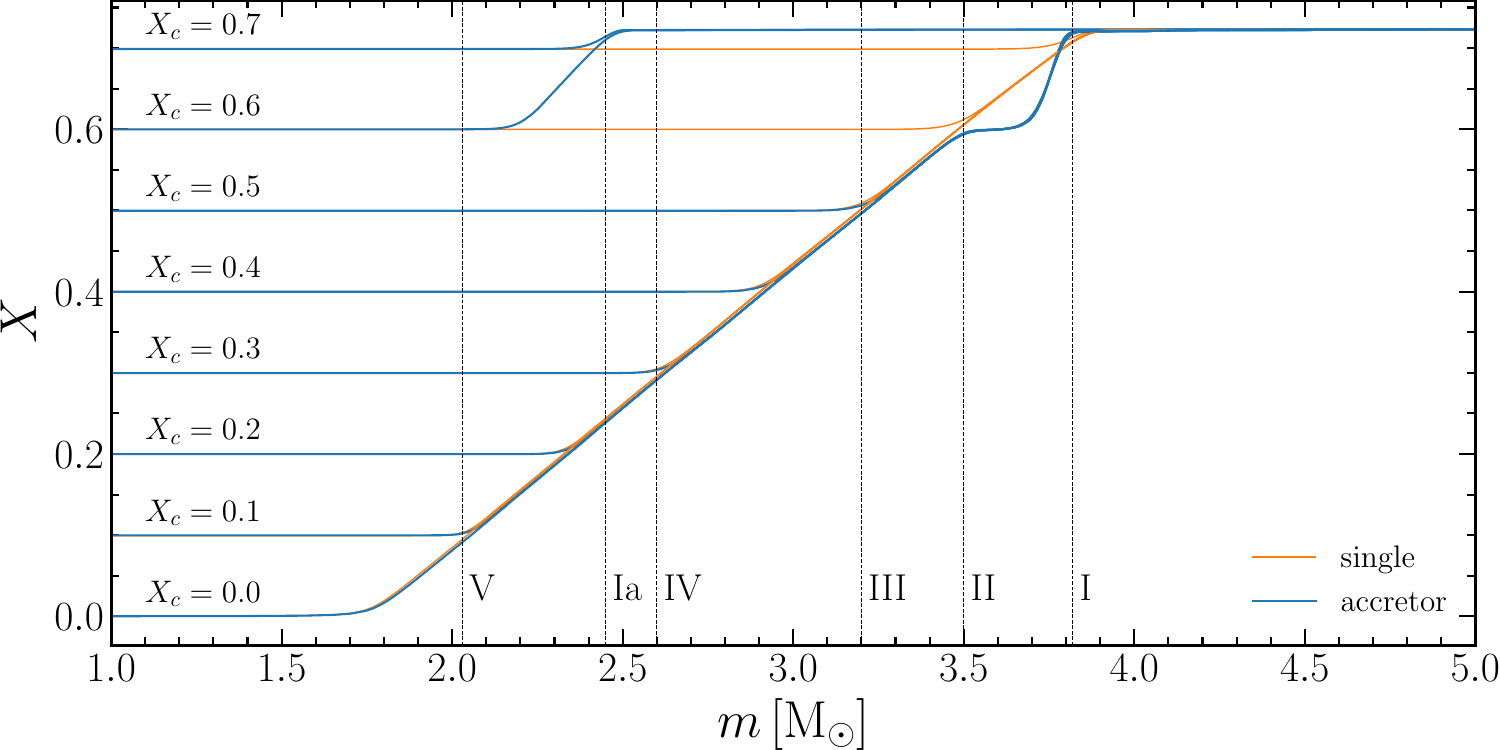}
    \caption{Hydrogen abundances $X$ in a function of a mass for different evolutionary stages of 10\Msun\ single (orange lines) and accretor (blue lines) models. The vertical lines mark features corresponding to Figure~\ref{fig:grad_prop_diagrams}. For clarity, we only show a fraction of the inner mass coordinate.}
\label{fig:X_abundance}
\end{figure}

The oCZs play a critical role in reshaping the hydrogen abundance profile. As they develop within chemically stratified zones, they homogenize the local composition by erasing sharp gradients. The mixing within the oCZ reduces the mean molecular weight gradient leading to nearly uniform hydrogen abundance profiles followed by new discontinuities at the oCZ boundaries. These discontinuities increase $\nabla_{\rm L}$ and can suppress convection locally, effectively splitting or terminating the oCZs. At the same time, the chemically stratified region continues to be compressed by the outward growth of the convective core during the rejuvenation. This compression can again steepen $\nabla_{\rm rad}$, allowing it to exceed $\nabla_{\rm L}$ and temporarily revive convection. 
During the mass transfer episode, multiple density bumps form at the boundaries of successive oCZs \citep{Miszuda2025_oCZs} altering the hydrogen abundance profiles. However, only the final feature, formed just before the disappearance of the last oCZ at the end of the mass-accretion phase, persists into the post-mass transfer phases. Because it forms sufficiently far from the convective core, it is not smoothed out by overshooting and thus remains as a lasting structural relic of the accretion history. The impact of this feature on the chemical structure is illustrated in \autoref{fig:X_abundance}, where we compare hydrogen abundance profiles at various evolutionary phases between the single-star (dashed lines) and accretor (solid lines) models. This abundance kink remains clearly visible for $X_c \leq 0.5$, for $3.5 < m/$\Msun\ $ < 3.8$, highlighting its persistence throughout the post-mass transfer main sequence evolution.

\begin{figure}[tbp]
    \centering
    \includegraphics[width=\linewidth]{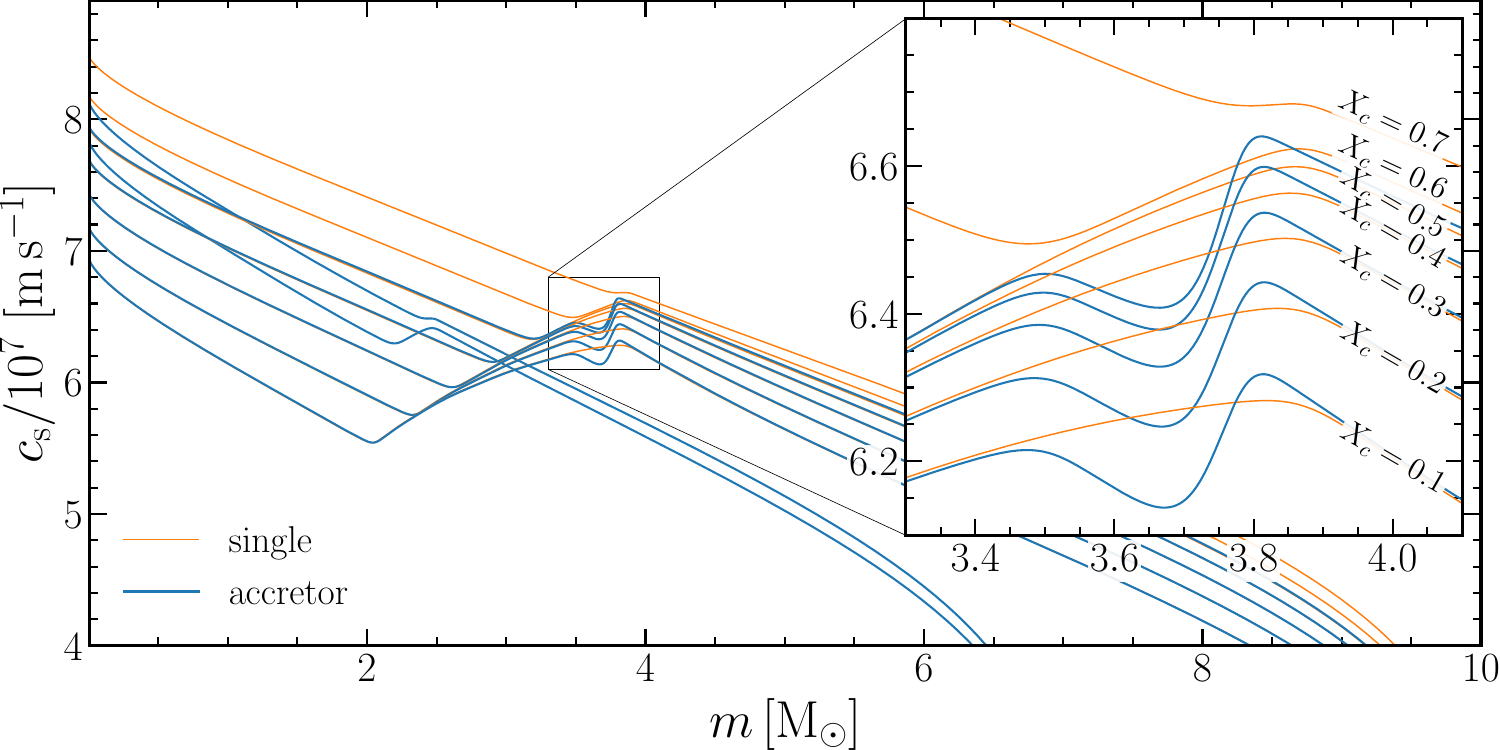}
    \caption{The sound speed $c_\mathrm{s}$ profiles at selected central hydrogen abundances $X_c$, for the single-star (orange lines) and the accretor (blue lines) models.}
\label{fig:c_sound}
\end{figure}

The structural imprint of the density bump is also reflected in the sound-speed profile. Since the adiabatic sound speed is defined as 
\begin{equation}
c_s^2 = \Gamma_1 P / \rho,
\end{equation}
where $\Gamma_1$ is the first adiabatic exponent, $P$ is the pressure and $\rho$ is the density, hence any local modulation in the density at fixed pressure alters the sound speed. In the region near the CEB, where pressure gradients are smooth due to hydrostatic equilibrium, the density bump causes a clear deviation in the $c_s$ profile, as shown in \autoref{fig:c_sound}. This modulation has observable consequences, as it affects the propagation of $p$ modes and may alter mode trapping behaviour \citep[e.g.,][]{Dziembowski1993, Miglio2008}. We investigate these effects in more detail in the following sections.

\subsection{Asteroseismic properties of the accretor model}

There are two characteristic frequencies that determine the behaviour of stellar oscillations. These are the Lamb frequency \citep{1917RSPSA..93..114L}:
\begin{equation}
S_{\ell}^2 = \frac{\ell(\ell+1)c_s^2}{r^2},
\end{equation}
and the Brunt-V\"{a}is\"{a}l\"{a} frequency \citep{1925SSFCP...2...19V,1927QJRMS..53...30B}:
\begin{equation}
N^2 = g \left( \frac{1}{\Gamma_1 P} \frac{\mathrm{d} P}{\mathrm{d}r} - \frac{1}{\rho}\frac{\mathrm{d} \rho}{\mathrm{d}r} \right)
\end{equation}
\citep[e.g.,][]{Aerts2010}. Here, $\ell$ is the spherical harmonic degree, $c_s$ is the adiabatic speed of sound, $g$ is the local gravitational acceleration and $r$ is the radial coordinate.

The propagation of modes is determined by the relation between the oscillation frequency $\omega$ and the characteristic frequencies $N$ and $S_\ell$\footnote{Here, presented in an angular form.}. Modes for which $\omega^2 < N^2$ and $\omega^2 < S_\ell^2$ propagate as gravity modes ($g$ modes), for which buoyancy acts as the restoring force. They are confined to radiative zones, where $N^2 > 0$, while convective zones act as evanescent barriers \citep[e.g.,][]{Cox1980,Aerts2010}.
In chemically homogeneous, non-rotating and non-magnetic stars, the $g$ modes with consecutive radial orders $n$ and the same spherical degree $\ell$ exhibit equally spaced periods $\Delta P$. According to the asymptotic properties of $g$ modes, valid for $\ell \ll n$, this spacing remains constant and is given by \citep{Tassoul1980}:
\begin{equation}
\Delta P_{\ell, \rm asym} = \frac{2 \pi^2}{\sqrt{\ell(\ell+1)}} \left( \int_{\mathcal{C}} N r^{-1} \mathrm{d}r \right)^{-1}, \label{math:dP}
\end{equation}
where the integration is performed over the $g$ mode propagation cavity $\mathcal{C}$ with respect to the radial coordinate. This expression shows that the period spacing is determined by the integral of the \BV frequency, and is therefore highly sensitive to the presence of chemical gradients and sharp structural features in the deep interior of the star. In case of the main sequence massive stars, with convective core and radiative envelope, the observed value of $\Delta P$ is directly related to the extent of the core \citep[e.g.,][]{2015A&A...580A..27M, 2016ApJ...823..130M,2019MNRAS.485.3248M,2018A&A...614A.128P,2021NatAs...5..715P}.

In contrast, modes with $\omega^2 > N^2$ and $\omega^2 > S_\ell^2$ propagate as pressure modes ($p$ modes), for which pressure acts as the restoring force. These propagate in the outer layers, where the sound speed and density is lower and the acoustic cavity is located. In massive main sequence stars like $\beta$\,Cephei variables, only low- to intermediate-order $p$ modes are observed. These modes of moderate radial order can penetrate into the deep stellar interior, making them valuable probes of the region near the convective core.
Because of this dependence, $p$ mode frequencies can provide a complementary asteroseismic diagnostic to $g$ mode periods, especially in stars that have undergone structural changes due to processes such as mass accretion. Changes in the internal $c_s$ profile, for instance due to convective core expansion, envelope compression, or compositional readjustment, can subtly shift the $p$ mode frequencies and thus leave observable signatures in the mode spectrum.

\subsubsection{The Brunt-V\"{a}is\"{a}l\"{a} frequency}

The study of the Brunt-V\"{a}is\"{a}l\"{a} frequency $N$ profile is essential to understand how mass accretion modifies the internal structure of massive stars. In particular, it allows us to trace the chemical and thermal stratification that governs the propagation of buoyancy modes within a star. 
Following the ideal gas plus radiation pressure for a fully ionised gas approximation, $N$ can be expressed as 
\citep[e.g.,][]{1989nos..book.....U}:
\begin{equation}
N^2 \approx \frac{g^2 \rho}{P} \left[\frac{4\beta-3}{\beta}(\nabla_{\rm ad} - \nabla) + \nabla_{\mu} \right],
\end{equation}
where $\beta = P_{\rm gas}/P$ is the ratio of gas pressure to total pressure.

The chemical gradient $\nabla_\mu$ is extremely important in asteroseismology of massive stars with convective cores. As the star evolves on the main sequence, the fully mixed core shrinks leaving a gradient of hydrogen (see \autoref{fig:X_abundance}) and helium abundance. This causes a positive contribution to $\nabla_{\mu}$ and hence, an emergent rise of $N$ \citep{Kippenhahn1990}. As the star evolves and the convective core recedes, the peak extends to the deeper parts of the star in a transition region adjacent to the core.
The chemical gradient acts stabilising on the convection \citep{Ledoux1947}, and thus this region defines a boundary for the $g$ mode cavity as the trapping occurs in the $\mu$ gradient zone \citep{Dziembowski1993}.

\begin{figure}[t]
    \centering
    \includegraphics[width=\linewidth]{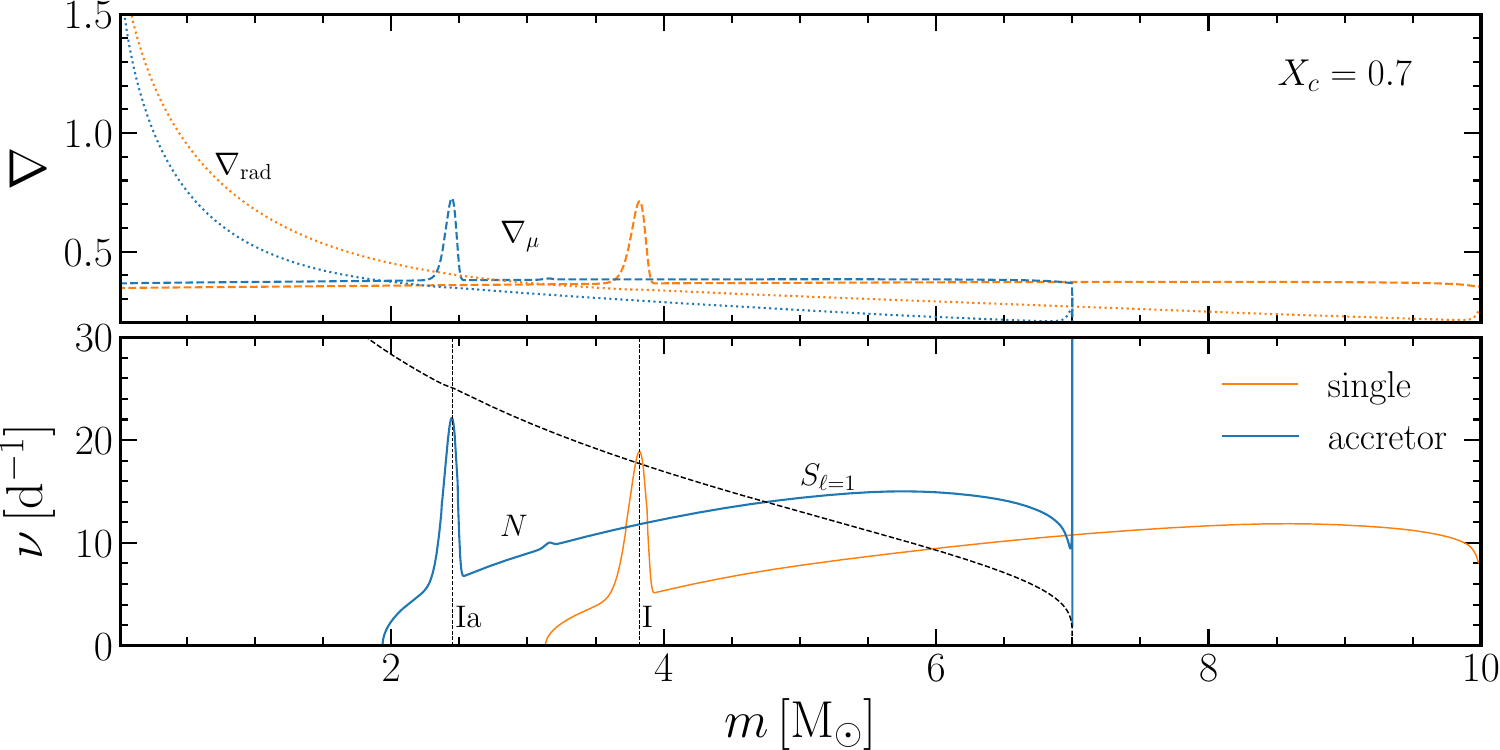} \\
    \includegraphics[width=\linewidth]{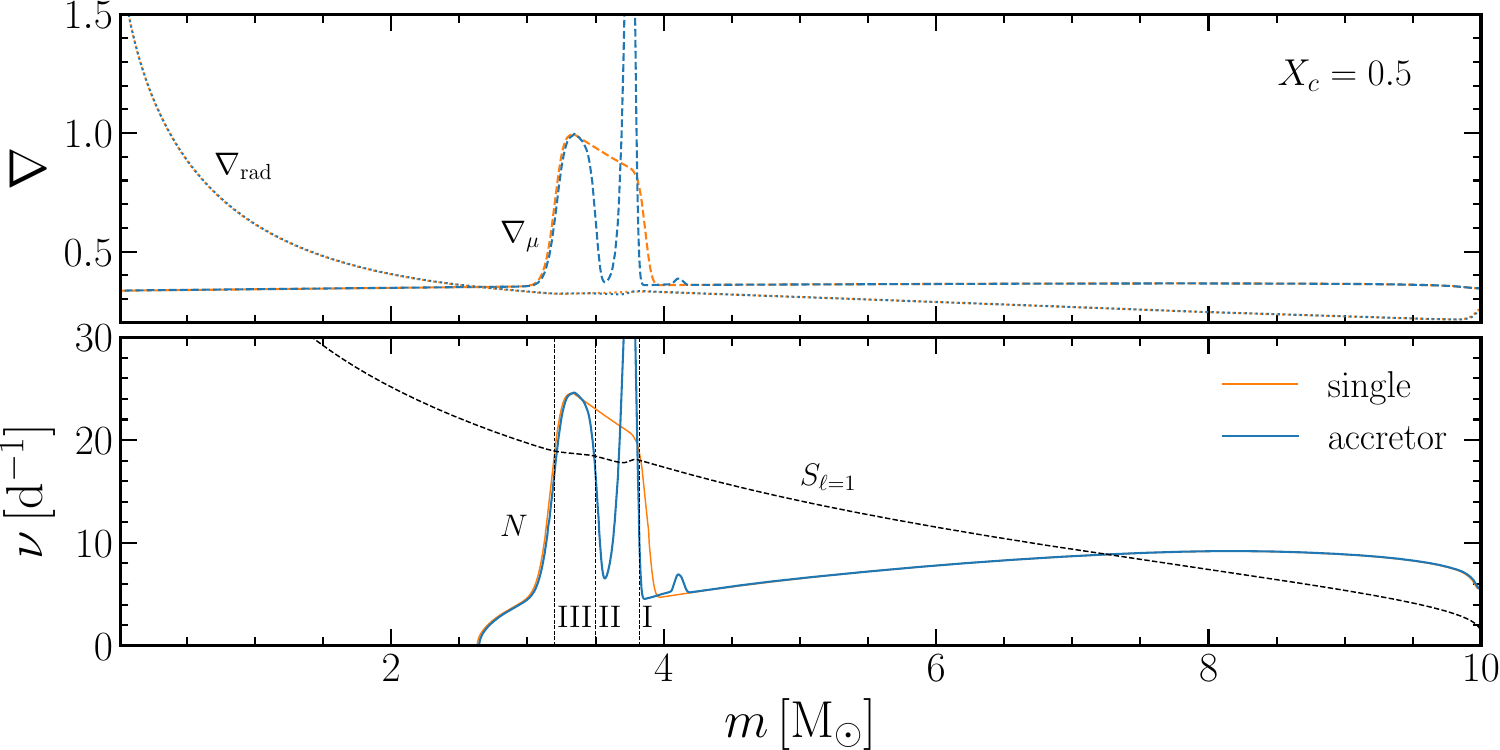} \\
    \includegraphics[width=\linewidth]{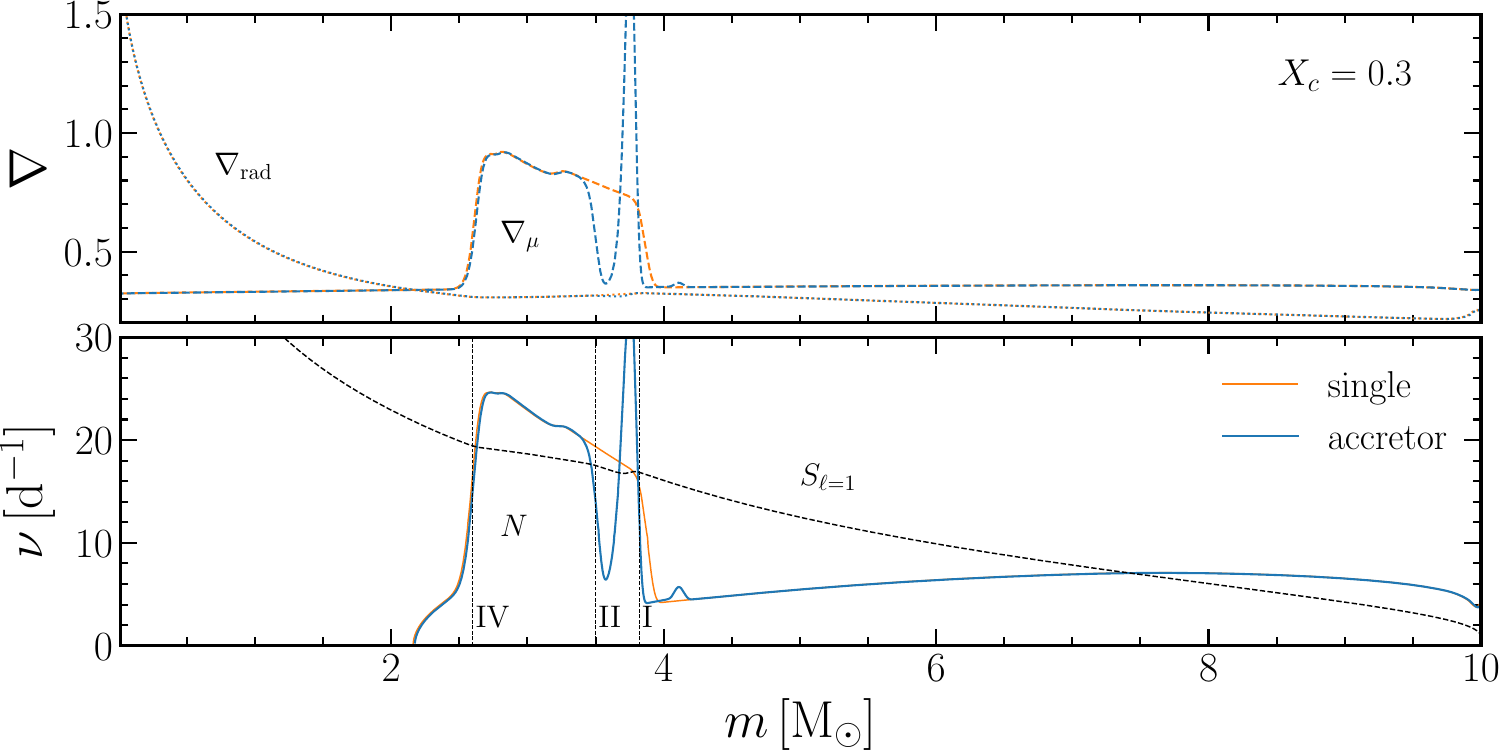} \\
    \includegraphics[width=\linewidth]{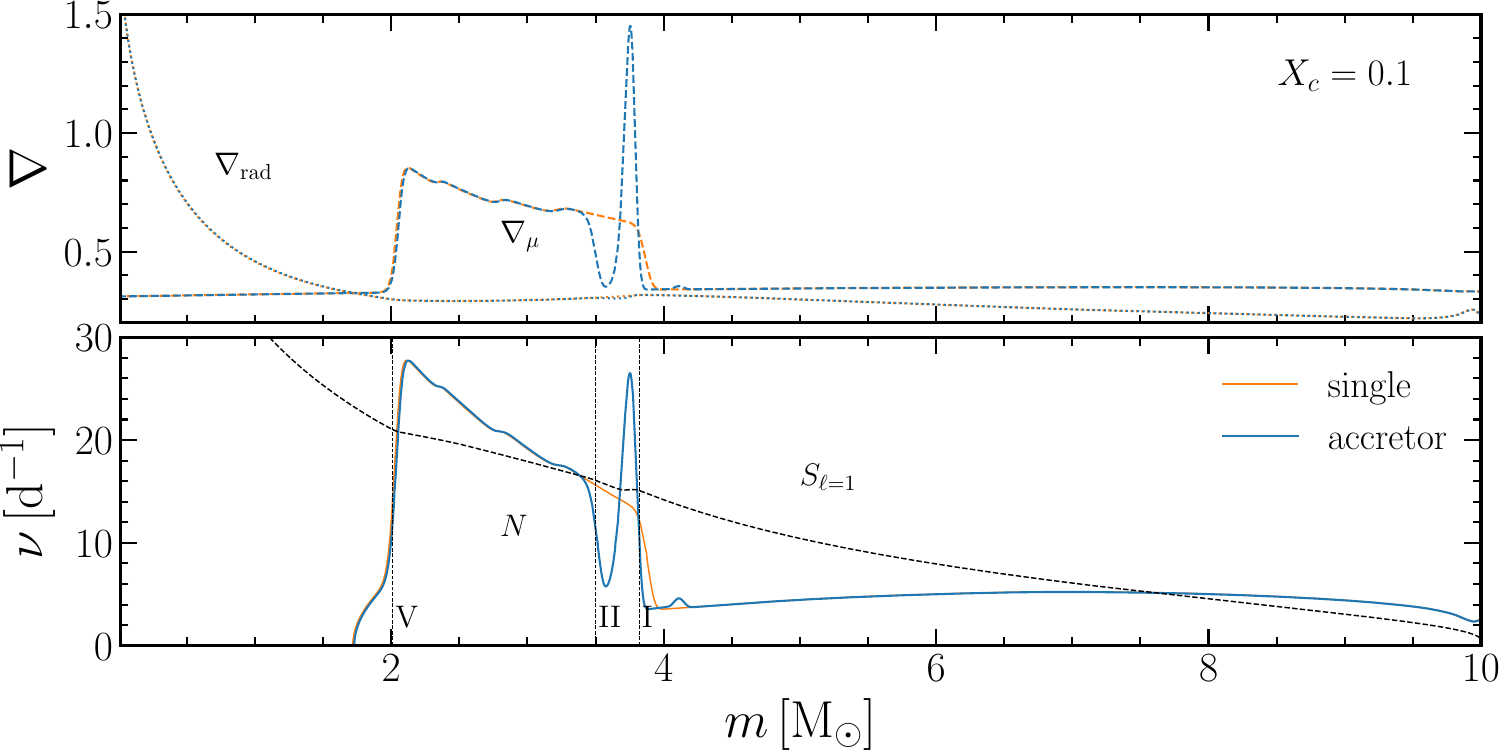} 
    \caption{The variation of radiative $\nabla_{\rm rad}$ (dotted lines) and chemical $\nabla_{\mu}$ gradients (dashed lines) on the top panels along with Lamb $S_{\ell=1}$ (dashed lines) and Brunt-V\"{a}is\"{a}l\"{a} $N$ frequencies (solid lines) on the bottom panels in a function of mass for 10\Msun\ single (orange lines) and accretor (blue lines) models. The vertical lines correspond to features shown in \autoref{fig:X_abundance}. We show results at four different evolutionary stages, $X_c=0.7, 0.5, 0.3$ and 0.1.  }
\label{fig:grad_prop_diagrams}
\end{figure}

By expansion of the convective core, the presence of the oCZs produce a deviation from the otherwise monotonic increase in hydrogen profile (see \autoref{fig:X_abundance}, solid lines) in the form of a plateau followed by an additional steep chemical gradient close to the outer boundary of the main chemical stratification. That gradient causes additional contribution to $\nabla_{\mu}$ (see top panels of \autoref{fig:grad_prop_diagrams}) and thus to $N$ (see lower panels of \autoref{fig:grad_prop_diagrams}) producing the bimodal distribution. As the star progresses through the main sequence, that feature persists, at least until TAMS, after which the star restructures its interior and develops a deep convective zone, that potentially can wash out this effect. The stability of the bimodal structure in $N$ across various evolutionary stages is visible in \autoref{fig:grad_prop_diagrams}.

\subsubsection{\texorpdfstring{$g$ modes}{g modes}}

The presence of two distinct peaks in the $N$ profile leads to the trapping of specific modes within specific stellar regions, thereby modifying their periods and inducing deviations from the asymptotic period spacing \citep[e.g.,][]{Dziembowski1993, Miglio2008,Wu2018}. These effects have been recently explored by \citet{Wagg2024} and \citet{2025A&A...698A..49H} for intermediate-mass slowly pulsating B star (SPB) pulsator models and massive merger models, respectively. We revisit them here, as we see a similar behaviour for the $g$ modes.

\begin{figure}[tbp]
    \centering
    \includegraphics[width=\linewidth]{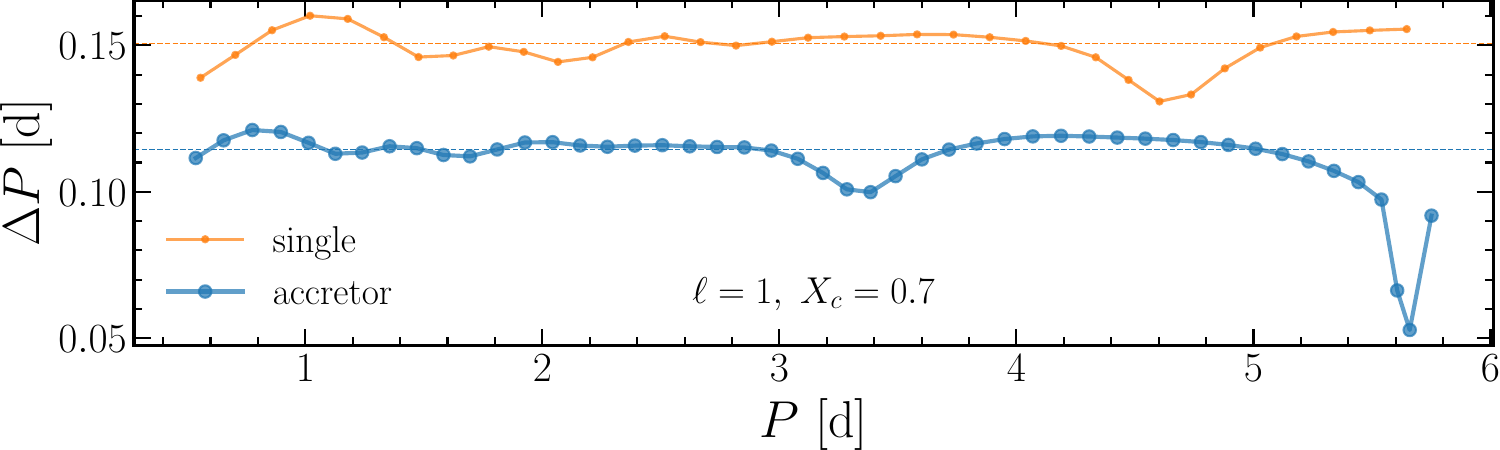} \\
    \includegraphics[width=\linewidth]{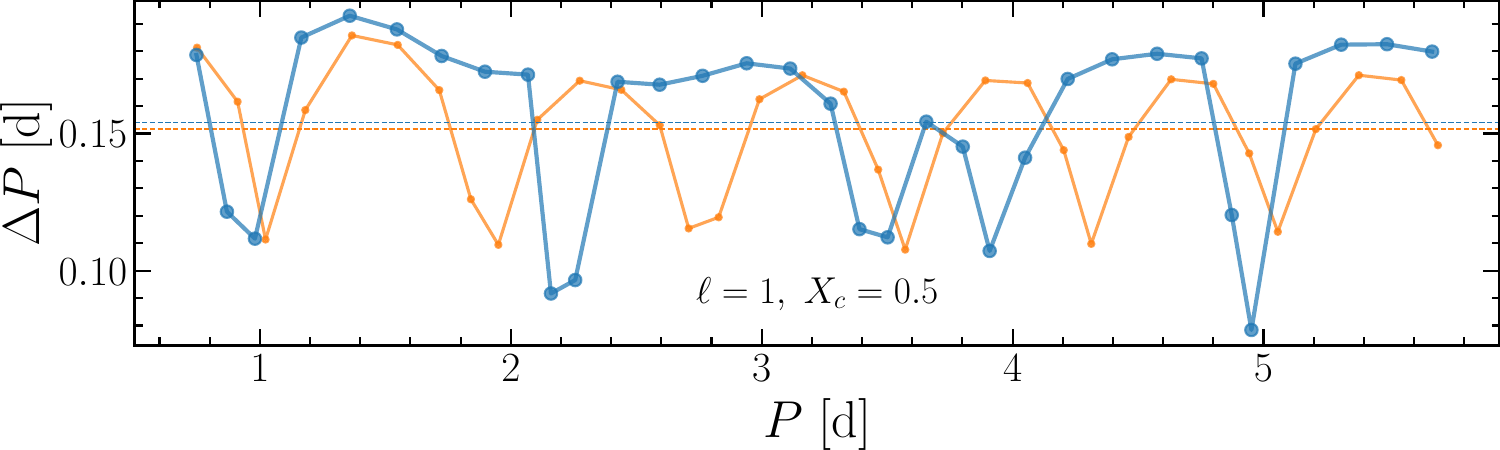} \\
    \includegraphics[width=\linewidth]{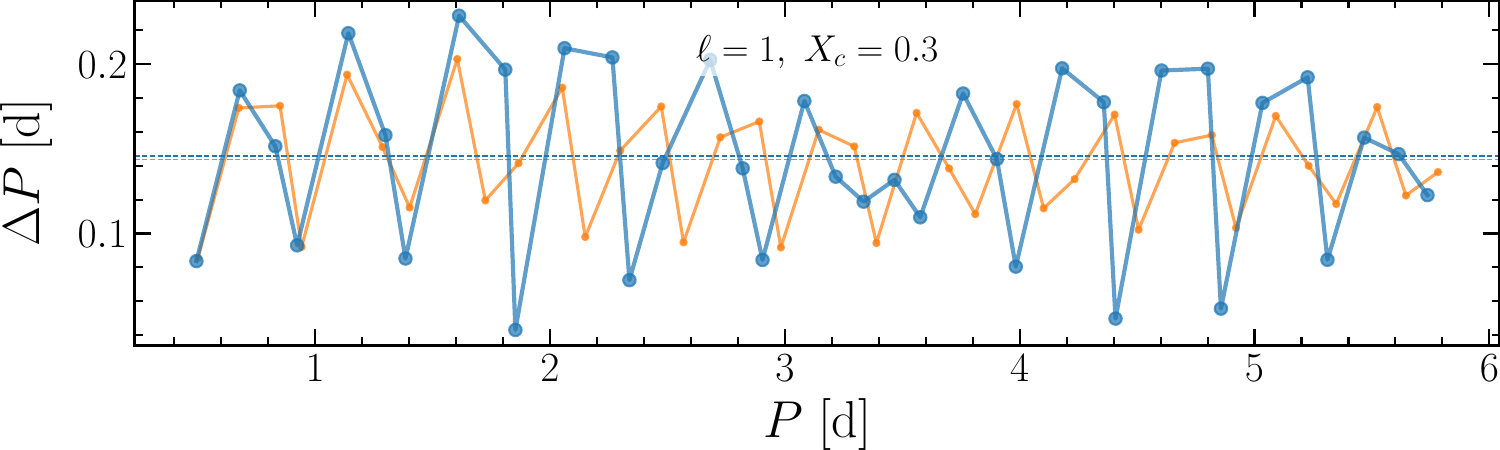} \\
    \includegraphics[width=\linewidth]{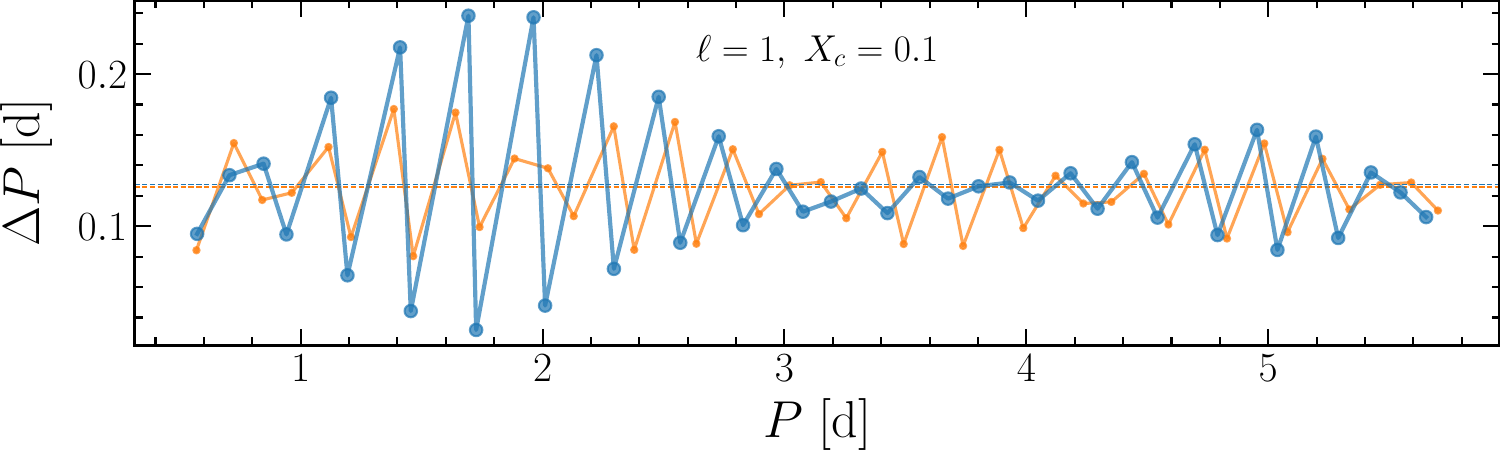} \\
    \includegraphics[width=\linewidth]{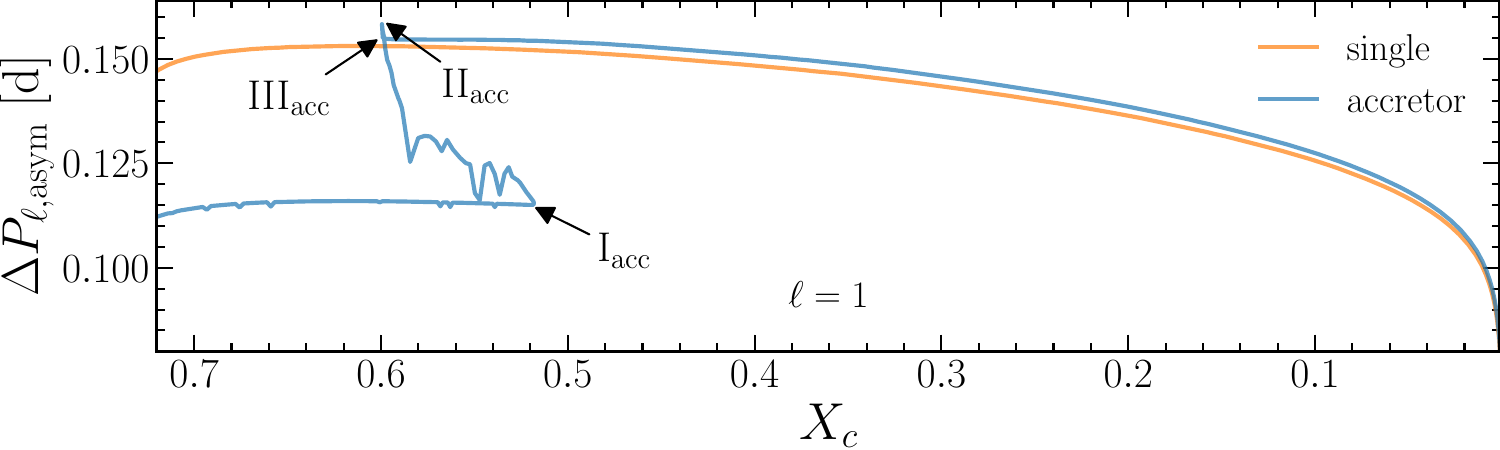} 
    \caption{The period spacing diagrams showing dipole $g$ modes, for both the accretor model (orange line) and the single-star model (blue line). The horizontal lines mark the asymptotic value of the period spacing. We show results for only four different evolutionary stages, $X_c=0.7, 0.5, 0.3$ and 0.1. The bottom panel shows the comparison of the $\ell = 1$ asymptotic period spacing evolution between the accretor and the single model. We mark points corresponding to key evolutionary changes of the accretor according to \autoref{fig:HR}.}
\label{fig:dP}
\end{figure}

As the accretor evolves on the early main sequence, the period spacing closely follows the asymptotic value. However, once the convective core begins to recede and a chemical gradient forms at its boundary, the $\Delta P(P)$ pattern develops regular oscillations around the asymptotic value, typically within a $P \sim 1$~day range. During the mass transfer episode, the spacing pattern becomes increasingly irregular, with no clear periodicity or coherence. This behaviour reflects rapid structural changes, the emergence of the off-centre convective zones, and the development of a highly structured \BV profile shaped by composition gradients and complex mixing. In the final stages of the main sequence, the period spacing pattern stabilises and regains regularity with a modulated oscillatory behaviour, and the amplitude alternating between increasing and decreasing in a quasi-periodic manner. The amplitude varies in a cyclic fashion, resembling a beating pattern. This behaviour likely results from mode trapping between multiple internal layers, shaped by persistent chemical gradients near the convective core boundary, and reflects the increasingly stratified structure of the stellar interior as the star approaches the terminal age main sequence. Such a behaviour can be seen in \autoref{fig:dP}, in which we plot the dipole mode period spacing $\Delta P$ for the accretor and single star models, corresponding to four separate evolutionary cases: pre-MT case at $X_c=0.7$, and post-MT at $X_c=0.5, 0.3$ and 0.1. With horizontal lines, we mark the asymptotic $\Delta P$ values (see \autoref{math:dP}), using solid lines for the accretor model and dashed lines for the equivalent single model. At $X_c=0.7$, both models exhibit nearly constant period spacings, with almost all modes lying close to the asymptotic value. This indicates a very small chemical composition gradient. Additionally, due to different initial masses, the asymptotic values themselves differ between the models, reflecting the dependence of the \BV integral on the stellar size and internal structure.
As evolution proceeds, after the mass transfer and mass equalization, the period spacing patterns begin to overlap, yet subtle differences persist between the models. Characteristic dips in the period spacing diagram start to emerge, associated with the development of a $\mu$ gradient region just outside the convective core. These dips are caused by mode trapping and result from the local increase in the \BV frequency. Their presence and depth provide valuable asteroseismic diagnostics of the internal chemical stratification and the evolutionary state.

At the bottom panel of \autoref{fig:dP}, we also show the evolution of the $\ell=1$ asymptotic $\Delta P_{\ell, \rm asym}$ values as a function of central hydrogen abundance $X_c$. For the accretor model, $\Delta P_{\ell, \rm asym}$ remains nearly constant at $\Delta P_{\ell, \rm asym} \sim0.115$\,d from the ZAMS up to $X_c \approx 0.5$. Following the onset of mass accretion ($\mathrm{I_{acc}}$, see \autoref{fig:HR}), $\Delta P_{\ell, \rm asym}$ increases rapidly, reaching a peak value of $\Delta P_{\ell, \rm asym} \sim 0.16$\,d at $X_c \approx 0.6$ ($\mathrm{II_{acc}}$) and later stabilizes at $\Delta P_{\ell, \rm asym} \sim 0.155$\,d ($\mathrm{III_{acc}}$). After mass transfer ends, the accretor evolves similarly to the single-star model, but with asymptotic $\Delta P_{\ell, \rm asym}$ values consistently higher by about 2\%. This offset can be linked directly to changes in the \BV frequency profile induced by mass accretion and structural readjustment.

We confirm the findings of \citet{Wagg2024}, that the period spacing pattern has a larger amplitude for the accretor compared to the single-star models. Additionally, for specific periods, the accretor’s pattern can be out of phase with the single star’s. This may be the most evident for $X_c=0.1$, in the second-to-last panel of \autoref{fig:dP}. The observed changes in period spacings between the accreting and single-star models indicate that mass transfer leaves detectable asteroseismic signatures. The magnitude of these differences, particularly in the $g$ mode regime, suggests that they could be within the observational capabilities of missions such as \textit{Kepler} \citep{Koch2010}, \textit{TESS} \citep{2015JATIS...1a4003R} or their successors, such as \textit{PLATO} \citep{2014ExA....38..249R}, provided that a sufficient number of consecutive modes can be identified.

\subsubsection{\texorpdfstring{$p$ modes}{p modes}}

\begin{figure}[thbp]
    \centering
    \includegraphics[width=\linewidth]{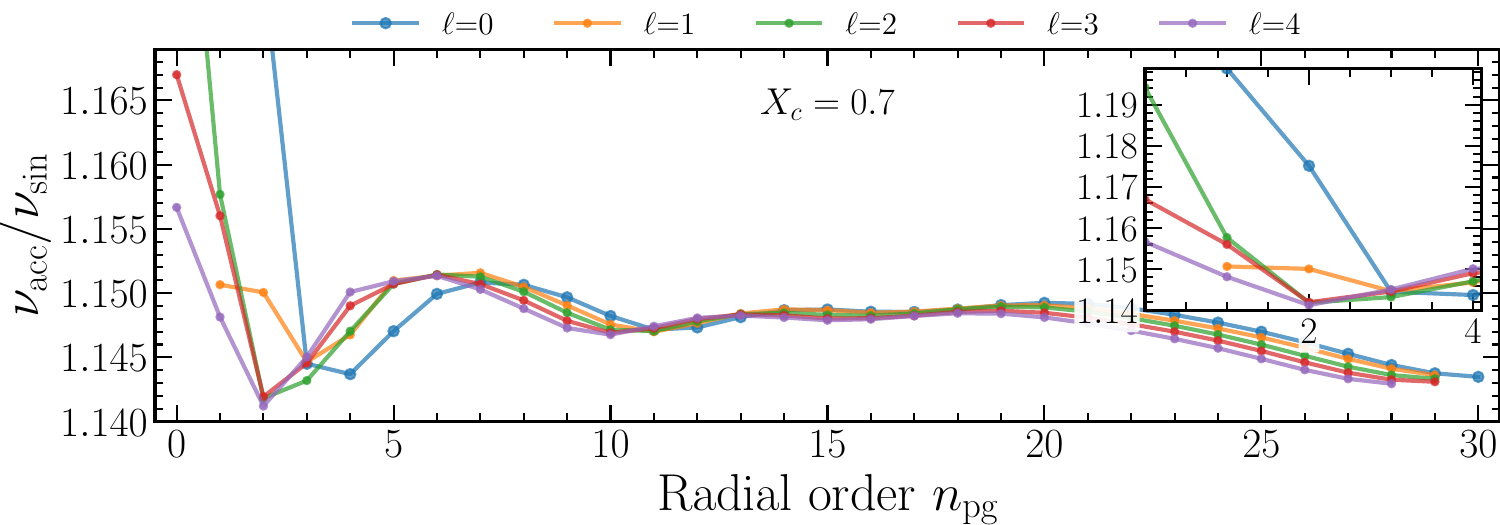} \\
    \includegraphics[width=\linewidth]{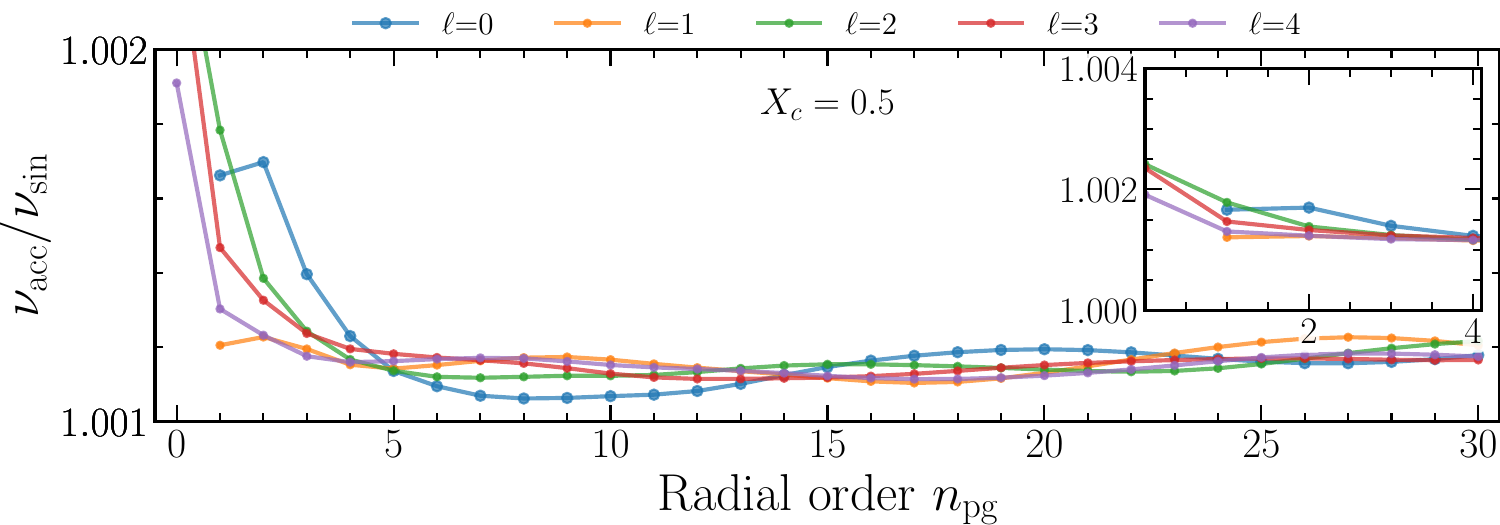} \\
    \includegraphics[width=\linewidth]{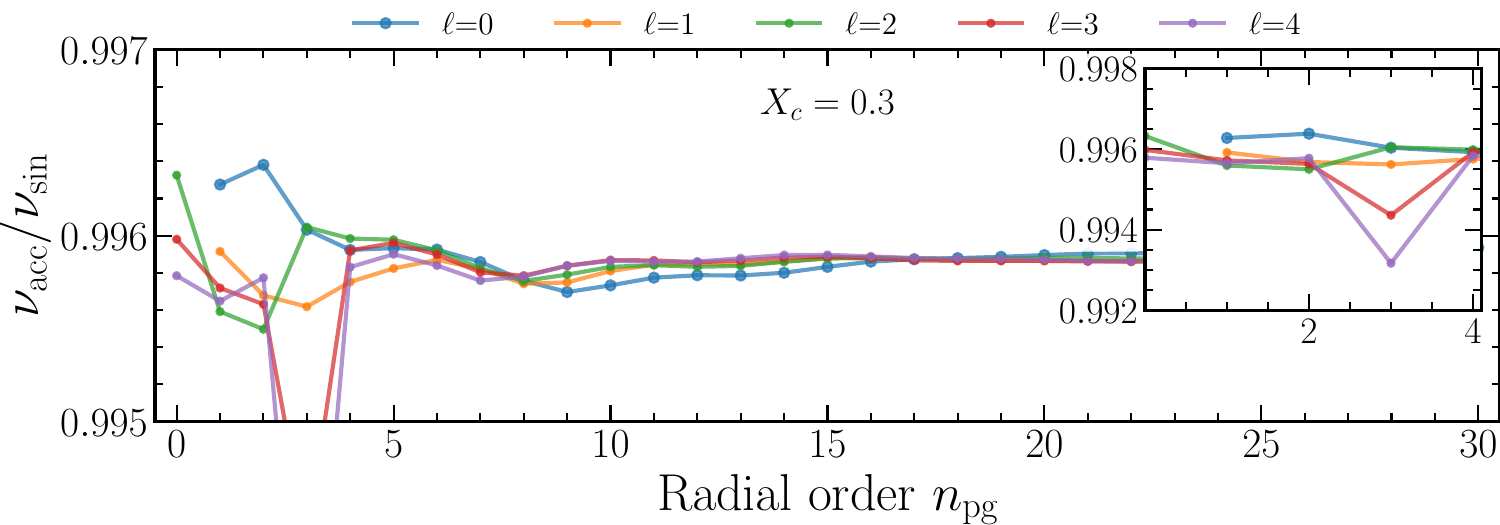} \\
    \includegraphics[width=\linewidth]{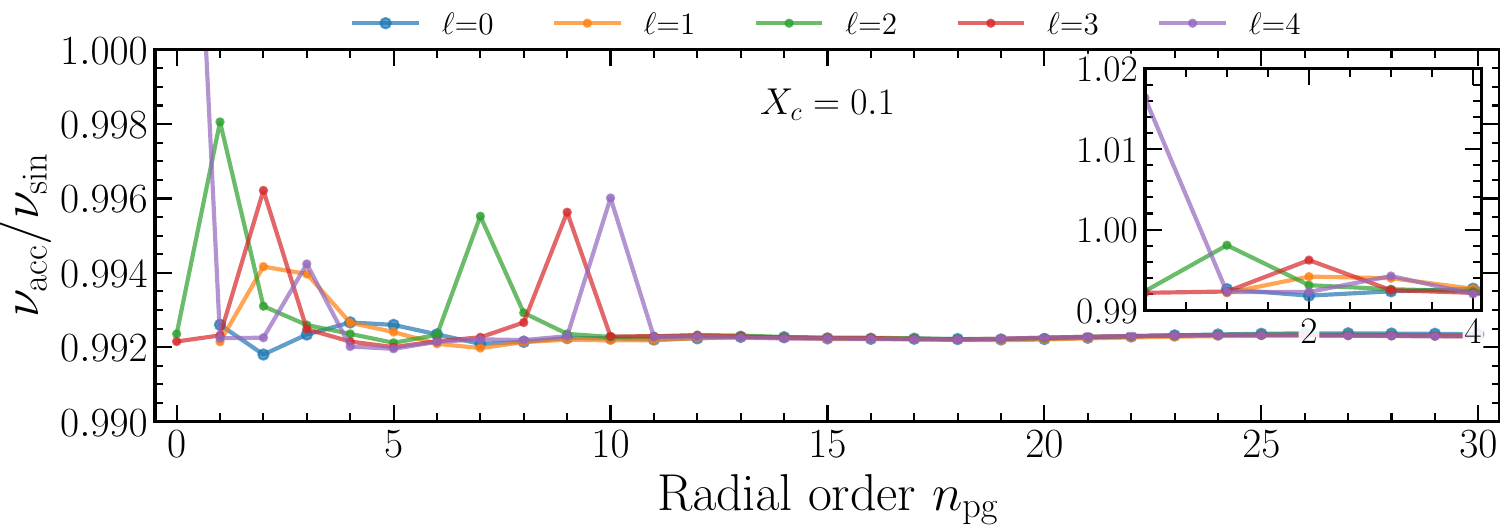}
    \caption{The accretor and single-star frequency ratios for a given modes at different evolutionary stages, $X_c=0.7, 0.5, 0.3$ and 0.1.}
\label{fig:freq_ratio}
\end{figure}

In the case of $p$ modes, we are primarily interested in low-degree, low to medium-order modes, since they are most likely to probe the deeper parts of the stellar interiors. For such modes, we calculate the accretor to single star frequency ratio $\nu_{\ell, n_{pg}}^{\rm acc}/\nu_{\ell, n_{pg}}^{\rm sin}$ for equal degree and order. We present the result in \autoref{fig:freq_ratio} for four separate evolutionary cases, $X_c=0.7, 0.5, 0.3$ and 0.1. Similarly as in \autoref{fig:dP}, the top panel shows the pre-MT case, where we have a mismatch between the stellar masses.  
However, as shown in the following panels, distinct deviations between accretor and single star frequencies appear in the low- to intermediate-order regime. These are superimposed on a systematic offset observed in the asymptotic region, where the frequency ratio $\nu_{\ell, n_{\rm pg}}^{\rm acc}/\nu_{\ell, n_{\rm pg}}^{\rm sin}$ reaches approximately 1.0012 for $X_c = 0.5$, 0.996 for $X_c = 0.3$, and 0.9925 for $X_c = 0.1$.
To illustrate this in more detail, we select the representative model at $X_c = 0.1$ and list the corresponding frequencies and their exact ratios in \autoref{tab:mode_freq}. As can be seen, the most affected frequencies are in the $5-15$\,\cpd\ range, which is the typical observed range for $\beta$\,Cep stars \citep[eg.,][]{2005ApJS..158..193S}. 
These non-asymptotic $p$ modes are more sensitive to conditions in the near-core region, including the sound-speed gradient and the local stratification shaped by previous mass accretion. As a result, their frequency deviations provide direct seismic evidence of internal restructuring. In order to quantify this sensitivity and link the frequency shifts to structural differences, we compute the weight functions. These allow us to identify the specific layers where the eigenfunctions contribute most significantly to the observed frequency deviations.

\begin{table}
    \centering
    \setlength{\tabcolsep}{3pt}
    \begin{tabular}{lrrc}
        \hline\hline
        Mode & Accretor model & Single model & Frequency ratio\T\\
        frequency & [\cpd] & [\cpd] & \B\\
        \hline
        $\nu_{\ell=1,n=2}$  & 5.2558  & 5.2866  & 0.9942\T\\
        $\nu_{\ell=1,n=3}$  & 5.9026  & 5.9385  & 0.9940 \\
        $\nu_{\ell=2,n=1}$  & 5.6151  & 5.6261  & 0.9980 \\
        $\nu_{\ell=2,n=7}$  & 11.4773 & 11.5289 & 0.9955 \\
        $\nu_{\ell=3,n=2}$  & 7.3094  & 7.3372  & 0.9962 \\
        $\nu_{\ell=3,n=9}$  & 13.8466 & 13.9073 & 0.9956 \\
        $\nu_{\ell=4,n=0}$  & 5.8409  & 5.7469  & 1.0164 \\
        $\nu_{\ell=4,n=3}$  & 8.7445  & 8.7952  & 0.9942\\
        $\nu_{\ell=4,n=10}$ & 15.3680 & 15.4295 & 0.9960\B\\
        \hline\hline
    \end{tabular}
    \caption{A comparison between accretor and single-star frequencies for given $p$ modes, corresponding to an $X_c=0.1$ model.}
    \label{tab:mode_freq}
\end{table}

It was shown by \cite{Kawaler1985} and later revised by \cite{Townsend2023} that the variational expression for the eigenfrequency of a mode can be written as
\begin{equation}
\sigma^2(\mathbf{y}) = \frac{\int_0^R \left[ \mathcal{C}(\mathbf{y},r) + \mathcal{N}(\mathbf{y},r) + \mathcal{G}(\mathbf{y},r) \right] \rho r^2 \mathrm{d}r}{\int_0^R \mathcal{T}(\mathbf{y},r)\rho r^2 \mathrm{d}r}, \label{math:sigma}
\end{equation}
where
\begin{align}
\mathcal{C}(\mathbf{y},r) & = g^2 \ell(\ell + 1) S_{\ell}^{-2}(y_2-y_3)^2, \label{math:C} \\
\mathcal{N}(\mathbf{y},r) & = r^2 N^2 y_1^2, \label{math:N}\\
\mathcal{G}(\mathbf{y},r) & = - \frac{gr}{U} \left[ y_4 + (\ell+1)y_3 \right]^2 \label{math:G}
\end{align}
are weight functions and $\mathbf{y}(r) = \Big\{ y_1(r), y_2(r), y_3(r), y_4(r)\Big\} = \Big\{ \xi_r/r, (\sigma^2r/g)(\xi_h/r), \Phi'/(gr), (1/g)d\Phi'/dr\Big\}$  is a vector of eigenfunctions introduced by \cite{Dziembowski1971}, corresponding to the radial displacement, the horizontal displacement, the gravitational potential perturbation and its derivative. Weight functions highlight the internal regions of the star that contribute to the mode’s frequency, revealing its sensitivity to specific layers and hence establishing its main formation regions. These have been used extensively to examine the mode trapping properties in subdwarf B stars \citep{Charpinet2000,Guyot2025}.

The function in the denominator,
\begin{equation}
\mathcal{T}(\mathbf{y},r) = r^2 \left[ y_1^2 + \ell(\ell+1) \left( \frac{g}{r \sigma^2} \right)^2 y_2^2\right],
\end{equation}
is proportional to the kinetic energy density and the $U$ quantity, appearing in \autoref{math:G}, is the homology invariant \citep[see eg.,][]{Kippenhahn1990}.

The frequency of a given mode $\nu_{\ell, n}$ can be expressed as 
\begin{equation}
\nu_{\ell,n}^2 = V (x=1),
\end{equation}
where $V$ is a cumulative integral of the weight functions $W$ over a normalised radius $x =r/R$,
\begin{equation}
V (x) = \int_0^x W(x') \mathrm{d}x'
\end{equation}
and
\begin{equation}
W(x) = \left( \frac{GM}{R^3} \right) \left( \frac{86400}{2\pi} \right)^2 \cdot \frac{\mathcal{C}(x) + \mathcal{N}(x) + \mathcal{G}(x)}{\int_0^R \mathcal{T}(r)\, \rho r^2 dr}
\end{equation}
The above is connected with \autoref{math:sigma} via
\begin{equation}
\nu = V(1)^{1/2} = \left( \frac{GM}{R^3} \right)^{1/2} \left( \frac{86\,400}{2\pi} \right)^{-1} \sigma.
\end{equation}

\begin{figure*}[hbt!]
  \centering
  \setlength{\tabcolsep}{1pt}
  \begin{tabular}{cc}
      \includegraphics[width=0.5\textwidth]{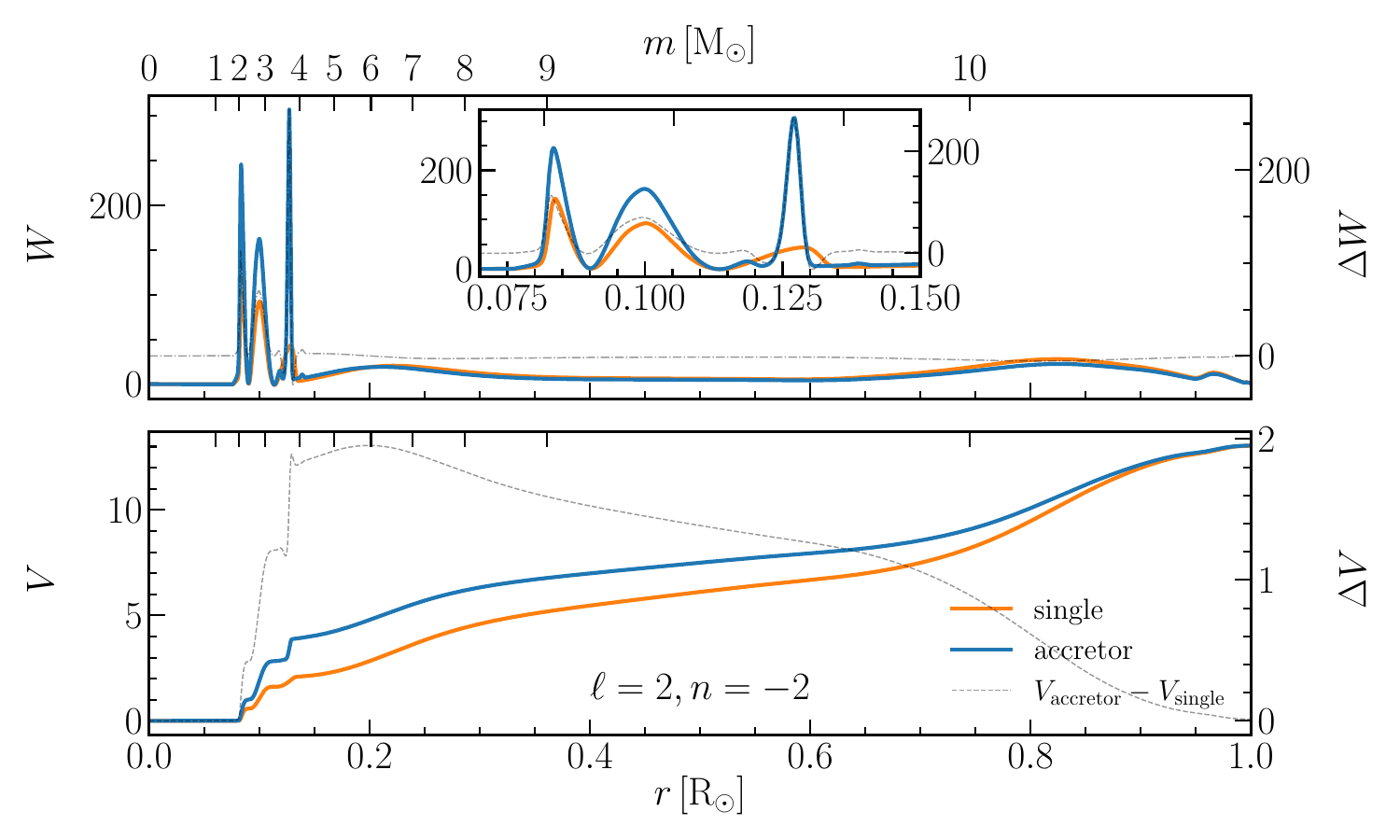} &
      \includegraphics[width=0.5\textwidth]{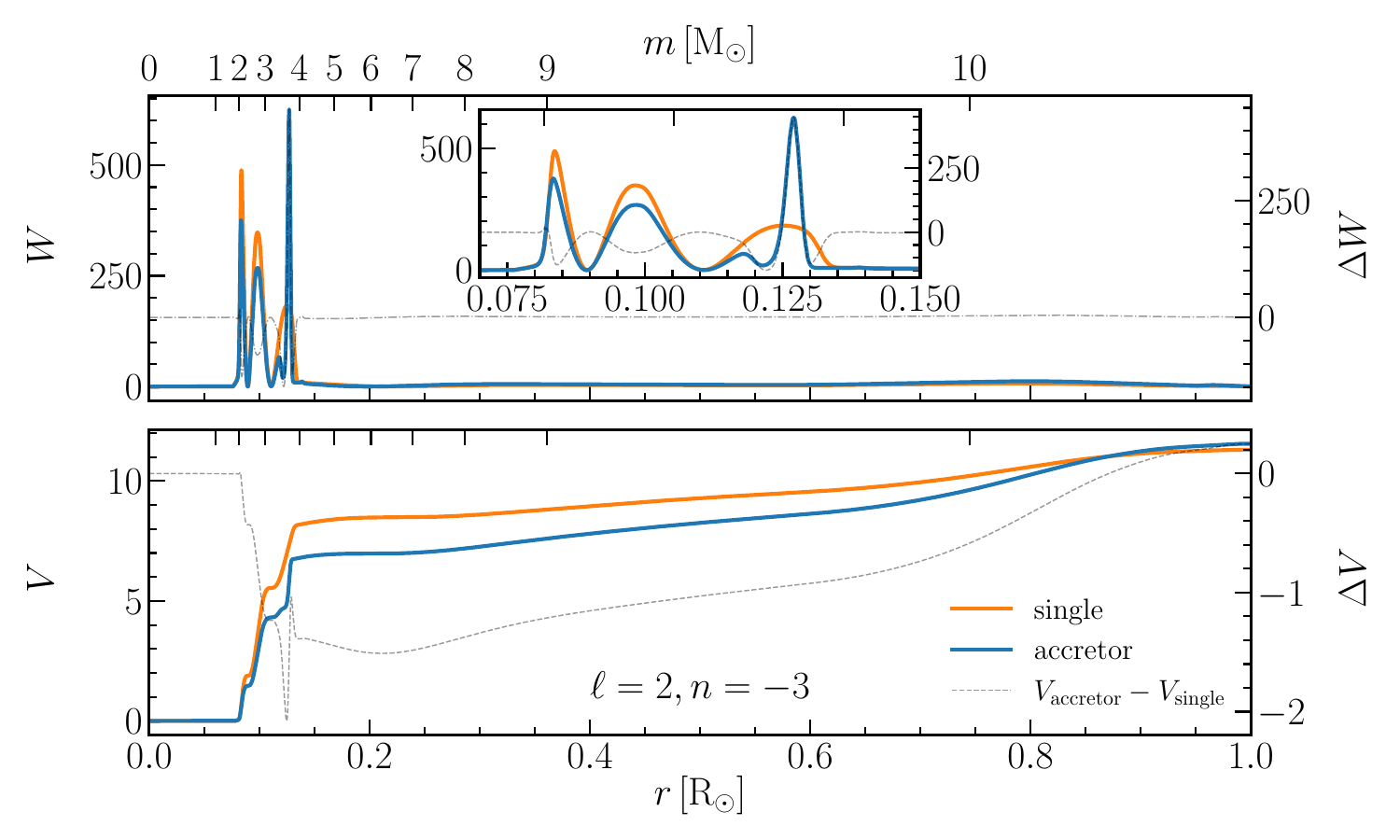} \\
  \end{tabular}

  \caption{Weight functions $W$ and their cumulative integrals $V$ for the specific $p$ modes (top panels) listed in \autoref{tab:mode_freq} at $X_c=0.1$ and for low-order $g$ modes, showing most distinct deviations between accretor (blue lines) and single-star (orange lines) frequencies.}
  \label{fig:weight_functions}
\end{figure*}

The weight functions help to identify the regions in the star to which a given mode is most sensitive, and their comparison allows us to understand how structural differences translate into changes in mode frequencies.
To investigate how the internal structure affects the oscillation frequencies, we examine the behaviour of the weight functions $W$ for selected modes. We focus on those with the largest frequency deviations between the two models at $X_c = 0.1$, namely the modes with $(\ell, n) = (2, 1)$ and $(\ell, n) = (4, 0)$ (see \autoref{fig:freq_ratio} and \autoref{tab:mode_freq}). Additionally, to illustrate the response of gravity modes to the internal structure, we show the weight functions for selected $g$ modes, i.e. $(\ell, n) = (2, -2)$ and $(\ell, n) = (2, -3)$. We deliberately select the low-order modes that can be trapped in the external peak region of the \BV frequency (see \autoref{fig:grad_prop_diagrams}), and are therefore particularly sensitive to the stratification region left by the oCZs. As a result, their character may differ significantly from that of the corresponding modes in single-star models, reflecting the structural imprint of the mass transfer history.
These weight functions are shown in \autoref{fig:weight_functions}, together with their cumulative integrals $V$. The insets zoom into the region $0.07 < r/$\Rsun\ $< 0.15$ (corresponding to $1 < m/$\Msun\ $< 4.5$).

\begin{figure*}[hbt!]
    \centering
    \setlength{\tabcolsep}{1pt}
    \begin{tabular}{cc}
        \includegraphics[width=0.5\textwidth]{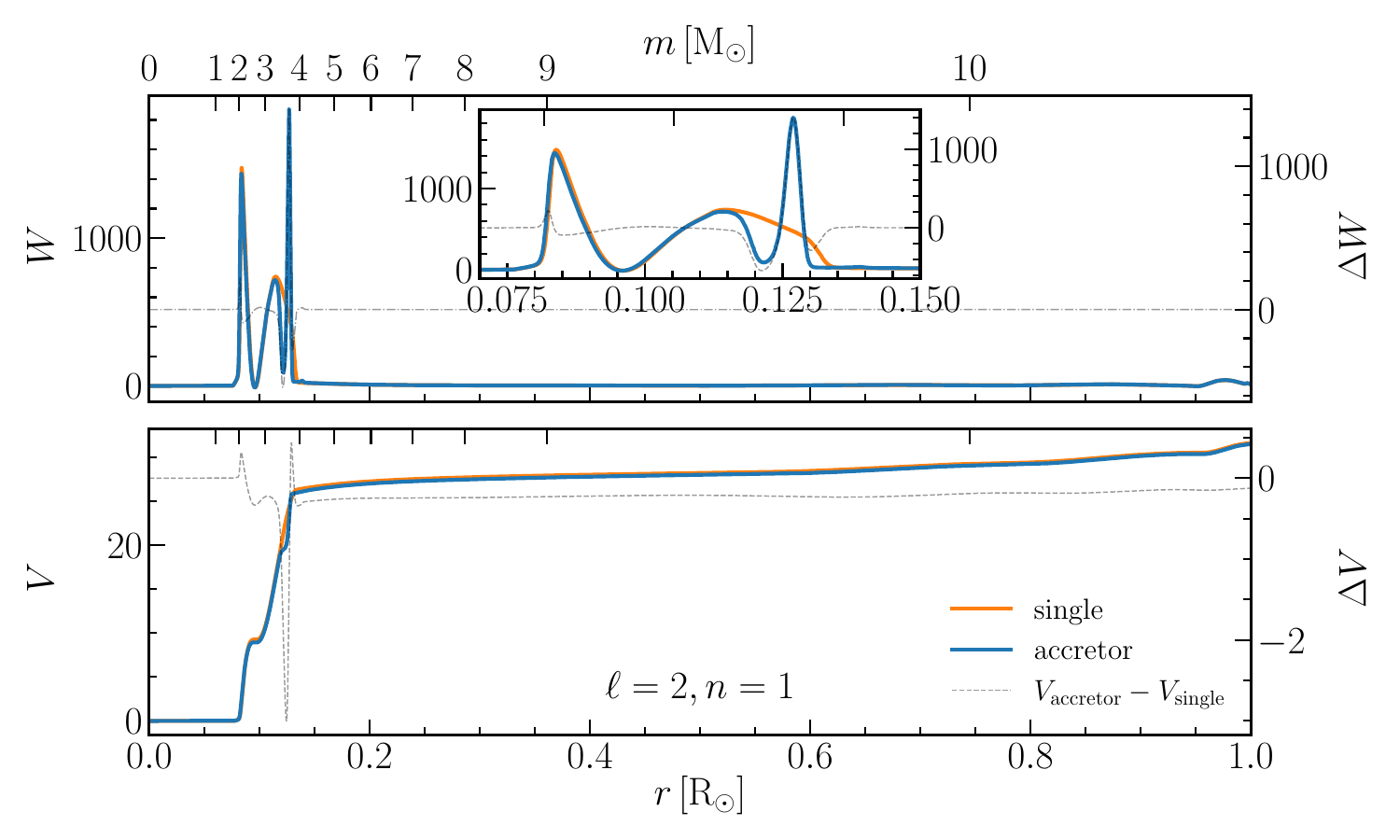} &
        \includegraphics[width=0.5\textwidth]{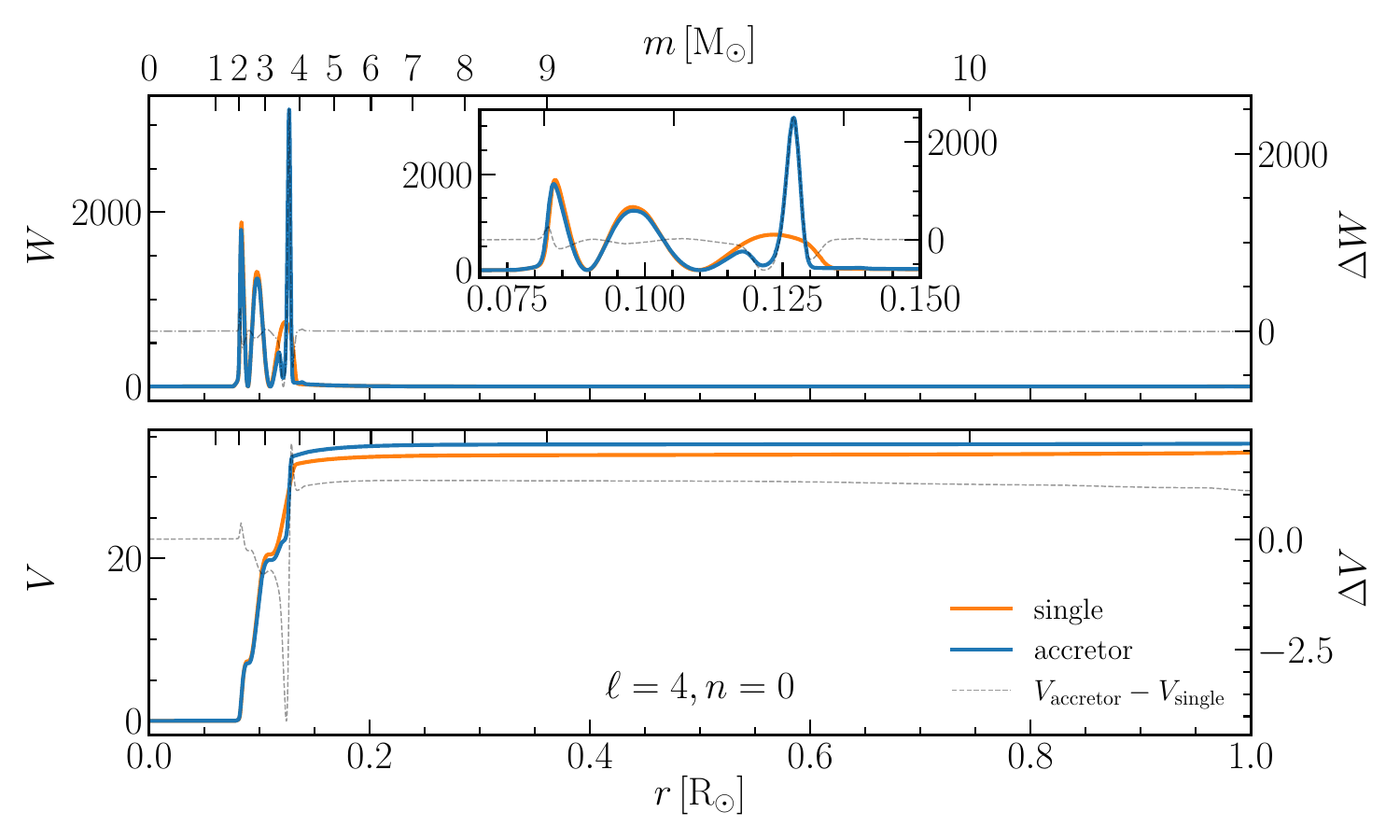} \\
    \end{tabular}
\end{figure*}

The analysis of $p$ mode frequency ratios further confirms that mass accretion subtly but measurably alters the sound-speed profile of the star. These variations, although smaller than those observed in the $g$ mode period spacings, offer complementary diagnostics sensitive to the layers close to the core and can thus provide additional constraints in asteroseismic modelling of mass-accreting systems. In particular, the differences in weight functions $\Delta W = W_{\rm accretor} - W_{\rm single}$ and the cumulative integrals $\Delta V = V_{\rm accretor} - V_{\rm single}$ indicate that the dominant contribution to the frequency shifts originates from the region $2 < m/$\Msun\ $ < 4$, which coincides with the zone of strongest divergence in the sound-speed profiles $c_s$ between the accretor and the single-star model (see \autoref{fig:c_sound}). The lower limit is, however, set by the size of the chemical gradient and thus dependent on the evolutionary phase we select to study. This region encompasses the outer edge of the convective core and adjacent stratified layers shaped by the accretion history. The frequency deviations can thus be interpreted as a seismic signature of the structural imprint left by the mass transfer event.
The magnitude of these $p$ mode frequency shifts, typically of the order of 0.5\%, correspond to absolute differences of $\sim 0.03 - 0.07$\cpd, in the $5 - 15$\,\cpd\ range. These are well above the typical observational uncertainties for \bcep\ stars \citep[$\sim 0.003$\%, e.g.,][]{2005MNRAS.360..619J,2023NatAs...7..913B}, and are therefore very likely to be detectable. However, as pointed out by \cite{Wagg2024}, we are still most likely to confuse the observed frequencies originating from binary evolution with those of a much younger single star. This degeneracy can lead to a misinterpretation of the star’s evolutionary state and age, potentially obscuring the effects of past mass transfer in asteroseismic modelling.

\subsubsection{\texorpdfstring{High-order $p$ modes}{High-order p modes}}

Even though high-order p modes are not observed in $\beta$ Cephei pulsators, we explore the asymptotic properties of the $p$ modes, as their asymptotic behaviour might offer a valuable theoretical framework for comparison.

It is well known that a local glitch such as a discontinuity in sound speed or density can induce periodic variations in the $p$ mode frequencies \citep{Gough1990}.
For solar-like oscillators, the bottom of convective envelop and the partial ionization zone of helium, have been inferred from high-frequency $p$ modes \citep{Roxburgh1994,Basu1994}.
The observables include the small frequency separation $\delta_{02}=\nu_{n,\ell}-\nu_{n-1,\ell+2}$ or the second-order frequency differences $\nu_{n-1,l}-2\nu_{n,l}+\nu_{n+1,l} $ \citep{White2011,Verma2014}. 

From the asymptotic analysis, the small separation over large separation ratio $\delta_{02}/\Delta\nu$ can be approximated as 
\begin{equation}
\delta_{02}/\Delta\nu \approx \frac{(2\ell+3)}{4\pi^2\nu}\left(-\int{\frac{\mathrm{d}c}{\mathrm{d}r}}\frac{\mathrm{d}r}{c} +\frac{c(R)}{R}\right).
\label{math:dnu}
\end{equation}
Thus, the imprints of mass transfer on the sound speed gradient near the core directly affect the magnitude of small separation. The above approximation has been refined and applied to red giants recently by \cite{Ong2025} and \cite{Reyes2025}.

\begin{figure}[ht]
    \includegraphics[width=\linewidth]{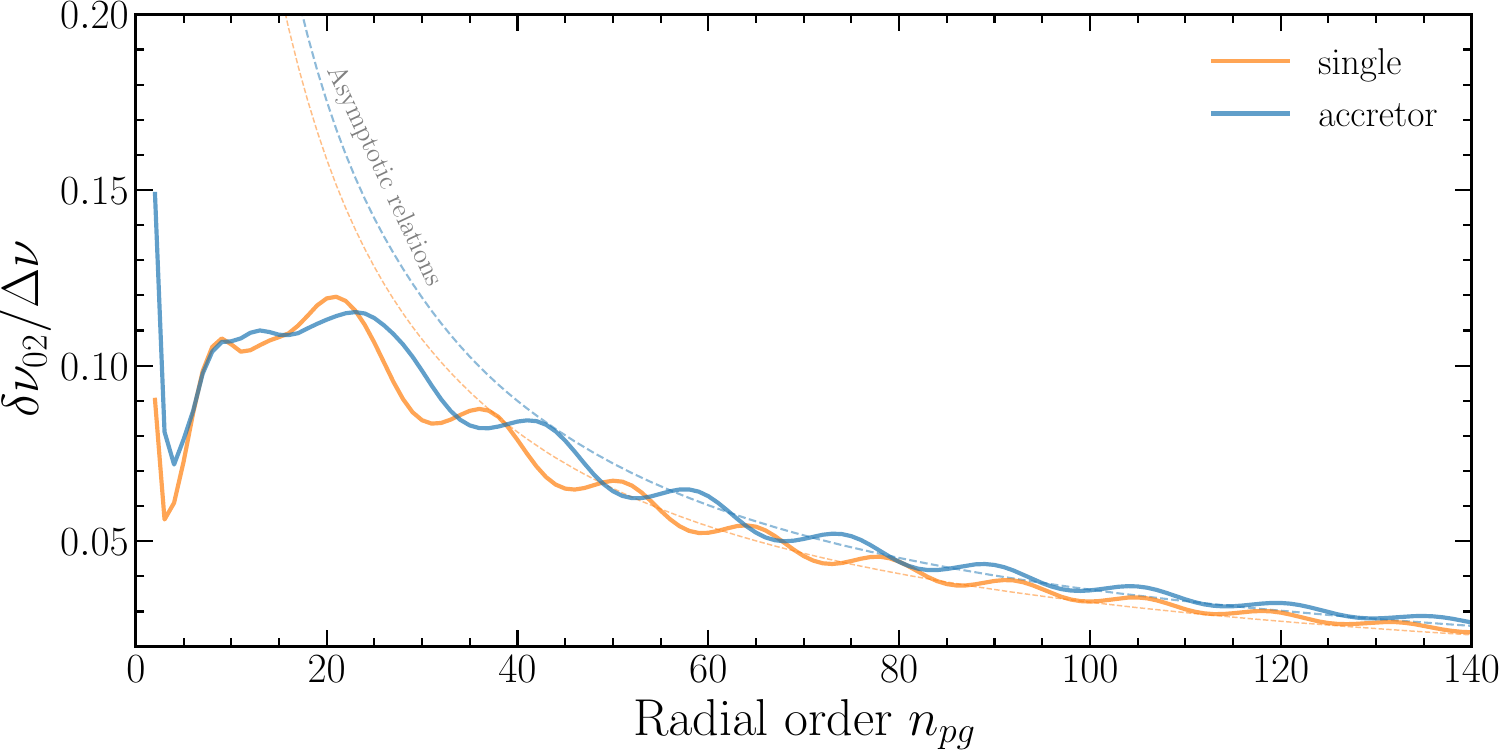}
    \caption{The ratio of the small frequency separation $\delta\nu_{02}$ derived from $\ell=0$ and $\ell=2$ $p$ modes, to the large frequency separation $\Delta\nu$, is shown as a function of radial order. The asymptotic values are represented by dashed lines, while the solid lines correspond to the results obtained from GYRE. Both the single-star (orange lines) and the accretor (blue lines) models are included for comparison.}
\label{fig:del02_ratio}
\end{figure}

\autoref{fig:del02_ratio} presents the ratio of the small to large frequency separations, computed using GYRE and from the leading-order asymptotic approximation for both the accretor and single-star models. As expected from theory, the asymptotic estimate of $\delta\nu_{02}/\Delta\nu$ decreases monotonically with increasing radial order $n_\mathrm{pg}$, a trend that aligns well with the GYRE calculations in the high-order regime. Notably, the accretor model exhibits a slightly small to larger separation ratio compared to the single-star model. This difference can be attributed to a larger integral of the sound-speed gradient in the accretor model (see \autoref{math:dnu}). Superimposed on the overall decline is a periodic modulation, which arises from the density discontinuity of the convective core boundary.

Following \citet{Roxburgh1994}, the amplitude of this variation depends on the magnitude of the density discontinuity, and the period of the variational component can be approximated by 
\begin{equation}
\Tilde{P}=(T/t_{\rm core})\Delta\nu,
\end{equation}
where $t$ is the acoustic radius 
\begin{equation}
t(r)=\int^{r}_0{\frac{\mathrm{d}r}{c}}, 
\end{equation}
and $T=t(r=R)$, $t_{\rm core}=t(r=r_{\rm core})$. Since the large frequency separation is given by $\Delta\nu \approx 1/(2T)$, we can estimate the frequency of the variation as 
\begin{equation}
\Tilde{\nu}=1/\Tilde{P}=2t_{\rm core}.
\end{equation}

We measure the periodic component in $\delta\nu_{02}/\Delta\nu$ by recording the local minima and calculating the averaged period from high radial order $p$ modes from GYRE ($n_\mathrm{pg} >30$) and find nice agreement with the expected frequency $2t_{\rm core}$. 
We note that $\delta\nu_{02}/\Delta\nu$ of the accretor model (solid lines) has a slightly larger period than that of the single model (dashed lines), indicating a larger $T/t_{\rm core}$ ratio. 

These results highlight the sensitivity of high-order $p$ modes to structural discontinuities introduced by mass transfer, and demonstrate how asymptotic diagnostics, even if not directly observable in $\beta$ Cephei stars, can provide valuable insights into the internal structure of post-interaction models.

\section{Summary} 
\label{sec:summary}

In this study, we investigated the structural and asteroseismic consequences of mass accretion onto a working example of a main sequence star with a final mass of $10\,\mathrm{M_\odot}$, representative of a typical $\beta$\,Cephei pulsator. We compare a single-star model and an accretor model that has undergone conservative mass transfer, focusing on key internal diagnostics such as density, chemical composition, sound speed, \BV frequency and $g$ mode and low-order $p$ mode frequencies.

Mass accretion significantly modifies the internal structure and asteroseismic properties of the $\beta$\,Cep accretor. The most prominent structural change is the formation of a local bump in the density profile just above the convective core. This feature arises during rejuvenation when the core expands and engulfs hydrogen-rich layers, creating a sharp $\mu$ gradient and a non-monotonic density bump. The bump persists throughout the main sequence evolution and corresponds to a lasting chemical discontinuity visible at the outer-boundary of the $\mu$ gradient region.

These structural modifications are reflected in the \BV frequency profiles. While the single-star model shows a smooth \BV profile shaped solely by core retreat and composition gradients, the accretor exhibits a bi-modal profile that results in additional mode trapping.

The $g$ mode period spacing diagram offers a sensitive asteroseismic probe of these structural changes. Before accretion, both models show near-constant $\Delta P$ values, closely matching the asymptotic expectation. However, during and after mass transfer, the period spacing of the accretor becomes irregular due to rapid internal restructuring. Once the star reaches post-accretion state, $\Delta P$ patterns regain coherence but develop characteristic modulations. These features result from mode trapping in the layered structure around the convective core and trace the evolving composition gradients. The quasi-periodic modulations in period spacing, including beating-like behaviour, serve as direct signatures of internal stratification shaped by mass accretion.

Low- and intermediate-order $p$ modes are likewise affected. Frequency ratios $\nu^{\rm acc}/\nu^{\rm sin}$ reveal systematic shifts up to $\sim 0.5$\%, corresponding to absolute differences of $0.03-0.07$\cpd\ in the observable $5-15$\cpd\ range for $\beta$\,Cephei stars. These shifts arise primarily from differences in the sound-speed profile near the core due to the local density bump. Weight function analysis confirms that these layers dominate the contribution to the observed frequency deviations.

Finally, we detect changes in the $\delta\nu_{02}/\Delta\nu$ patterns. The accretor shows a slightly longer modulation period, indicating a larger $T/t_{\rm core}$ ratio. This further confirms the seismic signature of a structurally more extended core boundary region caused by prior mass accretion.

In summary, we demonstrate that mass accretion leaves measurable imprints on the internal structure of mass-accreting stars, especially in density, chemical gradients, and sound-speed profiles. These translate into observable differences in both $g$ and $p$ mode pulsations. Period spacing modulations, mode trapping features, and systematic frequency shifts provide robust asteroseismic diagnostics of prior mass transfer events. Such signatures are within the sensitivity of current high-precision photometry, making asteroseismology a powerful tool to probe the evolutionary history of massive stars in binary systems. This study, although focused strictly on a high-mass stellar model, provides a general methodology and a clear prediction of the seismic signatures expected from mass accretion. The presented approach and results are supported by independent modelling efforts, such as those by \citet{Wagg2024} and \citet{2024A&A...690A..65H,2025A&A...698A..49H}, further reinforcing the diagnostic potential of pulsations for unveiling past binary interaction.

Future work should concentrate on the effect of the adopted physics, in particular rotation and the associated mixing, on the asteroseismic properties of post-mass-transfer stars. Differential rotation and enhanced mixing can substantially alter the internal chemical stratification and thus modify the driving and damping of pulsation modes, leading to different instability domains compared to single-star counterparts (see the discussion on mode driving in \citealt{Miszuda2021} and \citealt{Wagg2024}). This may directly impact which modes are preferentially excited and how their frequencies are distributed. In addition, employing machine-learning techniques to systematically identify discrepancies between dynamically inferred stellar parameters and those derived from seismic modelling could provide a powerful diagnostic tool to distinguish genuine binary interaction products from apparently single stars.

\begin{acknowledgements}
\noindent This work was supported by the (AM) Polish National Science Centre (NCN), grant number 2021/43/B/ST9/02972, (ZG) European Research Council (ERC) under the Horizon Europe programme (Synergy Grant agreement N◦101071505: 4D-STAR), and (RHDT) NASA grants 80NSSC24K0895 and 80NSSC23K1517. AM would like to thank G. Handler for his valuable comments on this manuscript.

Calculations have been carried out using resources provided by Wroc\l aw Centre for
Networking and Supercomputing (http://wcss.pl), grant no. 265. 

\end{acknowledgements}

\section*{Data Availability}
We make all files needed to recreate our \textsc{mesa-binary} and \textsc{gyre} results publicly available at Zenodo: \href{https://doi.org/10.5281/zenodo.15829740}{10.5281/zenodo.15829740}.

\section*{Software}
-\,\textsc{gyre} \citep{GYRE-Townsend2013,GYRE-Townsend2018,GYRE-Goldstein2020,GYRE-Sun2023}, \\
- \textsc{mesa} \citep{Paxton2011,Paxton2013,Paxton2015,Paxton2018,Paxton2019,Jermyn2023}, \\
- \textsc{pyMESAreader} (\href{https://billwolf.space/py\_mesa\_reader/index.html}{https://billwolf.space/py\_mesa\_reader/index.html}), \\
- \textsc{Python SciPy} \citep{2020SciPy-NMeth}

\bibliography{bibliography}
\bibliographystyle{aa}

\begin{appendix}
\section{\textsc{mesa} input physics}
\label{appendix:mesa}

The \textsc{mesa} code builds upon the efforts of many researchers who advanced our understanding of physics and relies on a variety of input microphysics data.
The \textsc{mesa} EOS is a blend of the OPAL \citep{Rogers2002}, SCVH \citep{Saumon1995}, FreeEOS \citep{Irwin2004}, HELM \citep{Timmes2000}, and PC \citep{Potekhin2010} EOSes.
Radiative opacities are primarily from the OPAL project \citep{Iglesias1993,Iglesias1996}, with data for lower temperatures from \citet{Ferguson2005} and data for  high temperatures, dominated by Compton-scattering from \citet{Buchler1976}. Electron conduction opacities are from \citet{Cassisi2007}.
Nuclear reaction rates are from JINA REACLIB \citep{Cyburt2010} plus additional tabulated weak reaction rates from \citet{Fuller1985}, \cite{Oda1994} and \cite{Langanke2000}. Screening is included via the prescription of \citet{Chugunov2007}. Thermal neutrino loss rates are from \citet{Itoh1996}.
The \textsc{mesa-binary} module allows for the construction of a binary model and the simultaneous evolution of its components, taking into account several important interactions between them. In particular, this module incorporates angular momentum evolution due to mass transfer.
Roche lobe radii in binary systems are computed using the fit of \citet{Eggleton1983}. mass transfer rates in Roche lobe overflowing binary systems are determined following the prescriptions of \citet{Ritter1988} and \citet{Kolb1990}.

\end{appendix}

\end{document}